\documentclass[journal]{IEEEtran}
\usepackage{array}
\usepackage{float}
\usepackage{booktabs}
\usepackage{multirow}
\usepackage{amsmath}
\usepackage{amsthm}
\usepackage{amssymb}
\usepackage{graphicx}

\makeatletter

\providecommand{\tabularnewline}{\\}
\floatstyle{ruled}
\newfloat{algorithm}{tbp}{loa}
\providecommand{\algorithmname}{Algorithm}
\floatname{algorithm}{\protect\algorithmname}
\usepackage{tikz}
\usetikzlibrary{calc}

\theoremstyle{plain}
\newtheorem{lem}{\protect\lemmaname}
\theoremstyle{plain}
\newtheorem{thm}{\protect\theoremname}

\IEEEoverridecommandlockouts
\usepackage{mathrsfs}
\usepackage{amsfonts}
\usepackage{ifpdf}
\usepackage{cite}

\usepackage{array}
\usepackage{mdwmath}
\usepackage{mdwtab}
\usepackage{eqparbox}
\usepackage{subfig}

\usepackage{algorithmic}
\usepackage{algorithm}
\usepackage{epsfig}
\usepackage{times}
\usepackage{threeparttable}
\usepackage{multirow}
\usepackage{xcolor}
\usepackage{amsfonts}\usepackage{graphicx}
\usepackage{textcomp}
\usepackage{xcolor}
\usepackage[normalem]{ulem}
\usepackage{tabularx}
\allowdisplaybreaks[4]

\def\BibTeX{{\rm B\kern-.05em{\sc i\kern-.025em b}\kern-.08em
		T\kern-.1667em\lower.7ex\hbox{E}\kern-.125emX}}

\usepackage{amsthm}

\@ifundefined{showcaptionsetup}{}{%
 \PassOptionsToPackage{caption=false}{subfig}}
\usepackage{subfig}
\makeatother

\providecommand{\lemmaname}{Lemma}
\providecommand{\theoremname}{Theorem}

\begin{document}
\title{Minimizing Task-Oriented Age of Information for Remote Monitoring
with Pre-Identification}
\author{Shuying~Gan, Xijun~Wang,~\IEEEmembership{Member,~IEEE,} Chao~Xu,~\IEEEmembership{Member,~IEEE,}
and Xiang~Chen,~\IEEEmembership{Member,~IEEE}\thanks{Part of this work was presented at the IEEE GLOBECOM, Dec. 2024 \cite{Gan2024}.}\thanks{S. Gan, X. Wang, and X. Chen are with School of Electronics and Information
Technology, Sun Yat-sen University, Guangzhou, 510006, China (e-mail:
ganshy7@mail2.sysu.edu.cn; wangxijun@mail.sysu.edu.cn; chenxiang@mail.sysu.edu.cn).}\thanks{C. Xu is with School of Information Engineering, Northwest A\&F University,
Yangling, 712100, China (e-mail: cxu@nwafu.edu.cn).}}
\maketitle
\begin{abstract}
The emergence of new intelligent applications has fostered the development
of a task-oriented communication paradigm, where a comprehensive,
universal, and practical metric is crucial for unleashing the potential
of this paradigm. To this end, we introduce an innovative metric,
the Task-oriented Age of Information (TAoI), to measure whether the
content of information is relevant to the system task, thereby assisting
the system in efficiently completing designated tasks.  We apply 
TAoI to a wireless monitoring system tasked with identifying targets
and transmitting their images for subsequent analysis. To minimize
TAoI and determine the optimal transmission policy, we formulate the
dynamic transmission problem as a Semi-Markov Decision Process (SMDP)
and transform it into an equivalent Markov Decision Process (MDP).
Our analysis demonstrates that the optimal policy is threshold-based
with respect to TAoI. Building on this, we propose a low-complexity
relative value iteration algorithm tailored to this threshold structure
to derive the optimal transmission policy. Additionally, we introduce
a simpler single-threshold policy, which, despite a slight performance
degradation, offers faster convergence. Comprehensive experiments
and simulations validate the superior performance of our optimal transmission
policy compared to two established baseline approaches.
\end{abstract}


\section{Introduction}

\label{Sec:Section 1}

In conventional communication, information is abstracted into bits
to achieve error-free bit replication from source to destination.
This approach has demonstrated tremendous success across various voice
and data communication systems \cite{Niu2022}. However, it primarily
focuses on the reliability of transmission at the physical layer,
often overlooking the semantic meaning and utility of the information
\cite{Getu2024}. With the rise of the intelligent Internet of Things
(IoT) era, communication systems are experiencing a paradigm shift,
transitioning from simple information reconstruction to directly supporting
complex intelligent operations, such as image recognition, object
detection, and industrial defect diagnosis.  To meet this challenge,
task-oriented communication has been proposed as a promising new communication
paradigm \cite{Shi2023}. By  selectively transmitting only the data
relevant to downstream tasks, this paradigm enables  systems to make
accurate inferences or decisions in the appropriate time and context.
 Consequently, its focus on essential data transmission significantly
reduces redundancy and enhances service quality \cite{Guenduez2023}.

Previous work focused on IoT devices with limited capabilities that
could only sense and transmit data, and on sampling or transmission
strategies to enhance system performance \cite{Ari2024,Shisher2023,Li2024,Holm2023}.
However, this paradigm has shifted significantly in recent years.
With the rapid advancement of hardware technologies, the computing
capabilities of IoT devices have significantly improved, paving the
way for task-oriented communication. Specifically, modern IoT devices
can preprocess data, including data compression, feature extraction,
and simple classification, which introduces greater flexibility and
intelligent potential to IoT systems \cite{Wang2022a,Sagduyu2023dl}.
This evolution necessitates leveraging these enhanced capabilities
for optimizing task-oriented communication in IoT systems. 

In evaluating task-oriented communication systems, researchers have
traditionally employed metrics that fall into two main categories:
content-based and timeliness-based. Content-based metrics, such as
Mean Squared Error (MSE) and accuracy, assess the fidelity and correctness
of information. MSE evaluates reconstruction tasks by measuring the
difference between original and reconstructed data, while accuracy
assesses classification quality based on the rate of correct identifications.
Optimizing MSE involves techniques like advanced encoding, error correction,
and physical-layer enhancements to ensure faithful data reconstruction
 \cite{Arafa2021,Weng2021,Hu2024}. Accuracy optimization focuses
on improving feature extraction, refining decision boundaries, and
optimizing receiver-side processing for enhanced recognition \cite{Shao2022,Shao2023,Sagduyu2024}.
However, these metrics primarily evaluate task-oriented communication
performance from a static perspective. 

Conversely, timeliness-based metrics, such as Age of Information (AoI),
focus on the freshness of information. AoI quantifies the time elapsed
since the generation of the last successfully received information
packet, crucial for real-time decision-making \cite{Yates2021}. AoI
optimization involves developing sampling strategies, scheduling methods,
and decision-making policies to maintain information freshness \cite{Sun2017,Xu2020,Zou2021}.
 While AoI is a pioneering metric evaluating communication system
performance from a temporal perspective, it does not inherently capture
the relevance of information content to specific downstream tasks.

In this work, we explore a remote monitoring system comprising an
IoT device and a receiver, where the objective is to monitor target
data for downstream task execution. Leveraging its computing capabilities,
the IoT device pre-identifies whether  captured data are related to
  downstream tasks and sends the pre-identification result to the
receiver. Upon receiving this potentially inaccurate result, the receiver
determines whether to request the IoT device to transmit the real-time
data based on the monitoring target. In contrast to our previous work
\cite{Gan2024}, this work considers an unreliable channel with transmission
failures and offers an in-depth analysis of the solution structure
for the optimal transmission policy. This analysis is crucial for
understanding how to effectively manage data transmissions over an
unreliable channel. The main contributions of this paper are as follows:
\begin{itemize}
\item We introduce a  novel task-oriented communication metric called Task-oriented
Age of Information (TAoI). TAoI is measured based on the success of
received information aligning with the desired target. Specifically,
if the successfully received information aligns with the desired target,
TAoI decreases; otherwise, it increases.  By integrating semantic
relevance and timeliness, TAoI enables the remote monitoring system
to efficiently transmit target data and effectively complete downstream
tasks.
\item We  study the transmission scheduling problem to minimize the long-term
average TAoI in the remote monitoring system.  Accounting for the
nonuniform duration of distinct actions, the  transmission scheduling
problem is formulated as an infinite-horizon Semi-Markov Decision
Process (SMDP). Subsequently, we transform it into an equivalent Markov
Decision Process (MDP) with uniform time steps. We prove that the
optimal transmission policy follows a threshold structure with respect
to the TAoI and propose a Relative Value Iteration (RVI) algorithm
with reduced computational complexity based on this structure. To
further accelerate convergence, a single-threshold policy is introduced,
which slightly underperforms the optimal policy in terms of TAoI.
\item We conduct experiments on the CIFAR-10 dataset to validate the feasibility
and effectiveness of the optimal transmission policy and the single-threshold
policy. In addition, we evaluate the performance of these two policies
through simulations and compare them with two baseline policies: the
always-transmit policy and the pre-identification based policy. The
results show that the optimal policy consistently achieves superior
performance,  while the single-threshold policy offers faster convergence
at the cost of some performance loss. Notably, when the target appearance
probability and the misidentification probabilities are all equal
to $0.5$, the optimal policy coincides with the single-threshold
policy. Moreover, when the transmission cost is $1$ or the target
appearance probability is $1$, the optimal policy reduces to the
always-transmit policy. When the misidentification probabilities are
$0$, it becomes the pre-identification based policy.  
\end{itemize}
The rest of this paper is organized as follows. Related work is reviewed
in Section II. Section III presents the system model and introduces
the proposed metric. In Section IV, we provide the SMDP formulation
of the problem, prove the threshold structure of the optimal transmission
policy, and propose a low-complexity RVI algorithm. Additionally,
a fast-converging single-threshold policy is introduced. Simulation
and experimental results are presented in Section V, followed by the
conclusion in Section VI.

\begin{figure*}[!t]
\centering\includegraphics[width=0.98\textwidth]{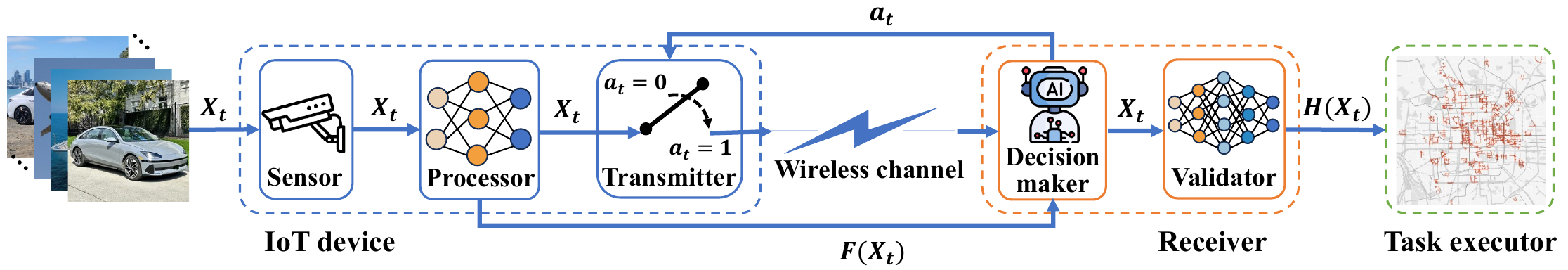}\caption{An illustration of the task-oriented monitoring system.}
\label{Fig:System_Model}
\end{figure*}

\section{Related Work}

\label{Sec:Section 2}

\subsection{Task-oriented IoT systems}

 Recent studies have increasingly focused on the optimization of task-oriented
communications in various IoT systems. In \cite{Ari2024}, the authors
studied a sampling and scheduling policy to minimize average inference
error for remote inference. Similarly, in \cite{Shisher2023}, the
authors aimed to minimize inference error by jointly optimizing feature
length and transmission scheduling with time-varying source data.
A system in which the semantic value of source data changes dynamically
was considered in \cite{Li2024}, where the authors investigated sampling
and decision-making strategies to maximize utility. The sensor scheduling
was investigated in \cite{Holm2023} for systems with multiple clients
with different, and potentially conflicting, objectives, where relevant
information is selected for a specific client.

In these studies \cite{Ari2024,Shisher2023,Li2024,Holm2023}, the
role of IoT devices is primarily limited to data sampling and transmission.
However, some research has begun to explore more sophisticated capabilities
at the device level. Specifically, a system with a preprocessing-enabled
IoT device was studied in \cite{Wang2022a}, with a focus on optimizing
the preprocessing and transmission processes. The transmitter-receiver
pair in the task-oriented IoT system was modeled as an encoder-decoder
pair in \cite{Sagduyu2023dl}, and jointly trained while considering
channel effects to enhance the performance of classification tasks.
These works extend the conventional IoT device by incorporating computational
capabilities, enabling on-device preprocessing to improve task performance.
However, the study in \cite{Wang2022a} provided a simplified preprocessing
mechanism without exploring the mapping between preprocessing and
downstream tasks, and \cite{Sagduyu2023dl} did not propose a comprehensive
optimization strategy for task-oriented IoT systems.

To overcome these limitations, this paper studies dynamic transmission
policies for task-oriented IoT systems with data preprocessing-enabled
IoT devices. This approach ensures improved task execution efficiency
while reducing redundant transmissions by explicitly considering the
relationship between preprocessing and downstream tasks, and developing
a corresponding optimization strategy. 

\subsection{Task-oriented Metrics }

Several recent studies have introduced various metrics based on  AoI
\cite{Zhong2018,Buyukates2022,Maatouk2020,Zheng2020,Wang2022}. In
\cite{Zhong2018}, the Age of Synchronization (AoS) was proposed to
measure the duration  that  information content at the receiver is
out of sync with the remote source. Similarly, the authors in \cite{Buyukates2022}
introduced the Age of Version (AoV), which calculates the number of
versions the receiver's information is outdated compared to the remote
source, where each source update  is considered a version change.
In \cite{Maatouk2020}, the authors proposed the Age of Incorrect
Information (AoII), which combines a time penalty  and an estimation
error penalty  to reflect the difference between the receiver's estimate
and the actual state of the system. The authors in \cite{Zheng2020}
introduced the Urgency of Information (UoI), defined as the product
of context-aware weight and the cost resulting from inaccurate status
estimation.  The age of changed information (AoCI) was proposed in
\cite{Wang2022}, which considers the impact of changes in information
content on the system.  Although these metrics  extend AoI by evaluating
information content from different perspectives, they do not directly
quantify the impact of information content on downstream tasks.

To address this gap, some new metrics have recently been designed
specifically to evaluate the impact of information content on downstream
tasks. In \cite{Kosta2020}, the authors introduced the Cost of Update
Delay (CoUD)  to describe the cost incurred by the destination due
to outdated information. Based on CoUD, they proposed a new metric
called Value of Information of Update (VoIU) to capture the importance
of received information, which is defined as the reduction in CoUD
when an update is received.   The Age of Loop (AoL) was designed in
\cite{Cao2023}, extending the uplink/downlink AoI into a closed-loop
AoI metric.  AoL  decreases only when the uplink status and downlink
command are successfully received.  The Age of Actuation (AoA), introduced
 in \cite{Nikkhah2023}, captures the elapsed time since the last
downstream task was performed at the destination based on information
received from the source. AoA  measures the timeliness of the receiver's
task execution using received information  and is a more general metric
than AoI. The authors in \cite{Fountoulakis2023} introduced the Cost
of Actuation Error (CoAE) to measure the cost incurred by errors in
downstream task execution due to inaccurate real-time estimation.
 However, these metrics do not fully reveal the relevance  between
the information content at the source and  the specific requirements
of the downstream task at the receiver Therefore, we propose a general
metric called TAoI for task-oriented communication that directly measures
whether the information content in the source is related to downstream
tasks.

\section{System Model}

\label{Sec:Section 3}

As shown in Fig. \ref{Fig:System_Model}, we consider a task-oriented
remote monitoring system consisting of an IoT device and  a receiver.
The IoT device includes a sensor, a processor, and a transmitter.
 The sensor in the IoT device captures real-time data, which are then
pre-identified by a lightweight binary classifier in the processor.
This classifier performs a preliminary identification of the data
(e.g., determining the presence of vehicles in images for highway
traffic analysis). The pre-identification result is then sent to 
the receiver. The receiver consists of a decision maker and a validator.
The decision maker utilizes the potentially inaccurate pre-identification
result  to determine whether to request the IoT device to send the
original data. Based on the decision maker's request, the transmitter
sends the data over an unreliable wireless channel.  Upon receiving
the data, the validator, equipped with a more complex and accurate
binary classifier, evaluates whether the data match the target criteria.
The validator then provides the validated data to a downstream task
executor. 

\begin{figure}[!t]
\centering\includegraphics[width=0.48\textwidth]{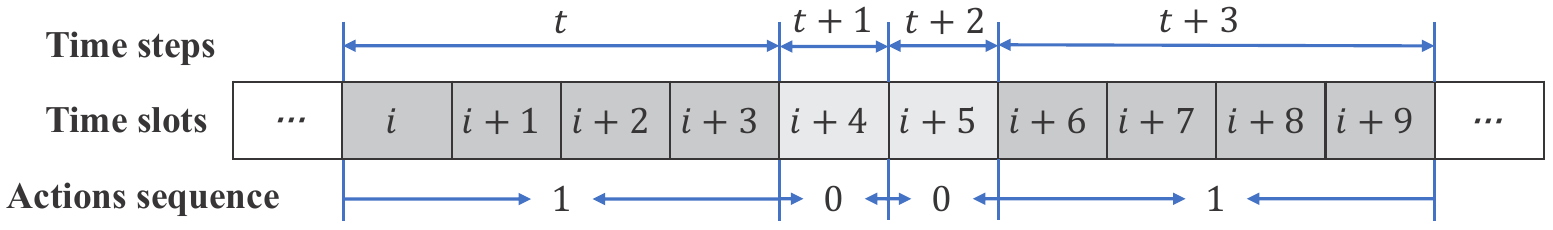}\caption{An illustration of time slots and time steps.}
\label{Fig:time_slot}
\end{figure}

 As shown in Fig. \ref{Fig:time_slot}, time is slotted,   with each
slot having a duration of  $\tau$.   We define a decision epoch for
the decision maker  as a time step. At the beginning of time step
$t$, the sensor captures  fresh data $X_{t}\in\mathcal{X}$. A relevance
indicator, denoted by $Y_{t}\in\{0,1\}$, indicates whether the data
contains content relevant to the task. Specifically, $Y_{t}=1$ indicates
that  $X_{t}$ contains content of interest to the downstream task,
and $Y_{t}=0$ otherwise. The probability of  $Y_{t}=1$ is $\mathrm{Pr}\left(Y_{t}=1\right)=q$,
and the probability  of $Y_{t}=0$  is $\mathrm{Pr}\left(Y_{t}=0\right)=1-q$.
Since the relevance indicator is initially unknown, the processor
pre-identifies the data $X_{t}$ and provides a pre-identification
result $F(X_{t})\in\{0,1\}$ as a potential noisy estimate of $Y_{t}$.
The processor then sends the obtained pre-identification result $F(X_{t})$
to the decision maker, where the transmission is assumed to be instantaneous
and error-free. However, The pre-identification itself may by inaccurate.
We define the misidentification probabilities as:
\begin{align}
 & p_{A}\triangleq\mathrm{Pr}(F(X_{t})=1|Y_{t}=0),\forall t,\label{Eq:misclassification_0}\\
 & p_{B}\triangleq\mathrm{Pr}(F(X_{t})=0|Y_{t}=1),\forall t.\label{Eq:misclassification_1}
\end{align}

\noindent Then,  the probability that the  pre-identification result
for data $X_{t}$ is 1 can be obtained as follows:
\begin{align}
g\triangleq & \mathrm{Pr}(F(X_{t})=1)=\mathrm{Pr}(F(X_{t})=1|Y_{t}=0)\mathrm{Pr}(Y_{t}=0)\nonumber \\
 & +\mathrm{Pr}(F(X_{t})=1|Y_{t}=1)\mathrm{Pr}(Y_{t}=1)\nonumber \\
= & {p}_{A}(1-q)+(1-{p}_{B})q,\forall t.\label{Eq:pre-identify}
\end{align}

Based on the pre-identification result $F(X_{t})$, the decision maker
determines whether to request the transmitter to transmit  $X_{t}$.
Let $a_{t}\in\left\{ 0,1\right\} $ denote the transmission decision
at time step $t$, where $a_{t}=1$ indicates that the transmitter
transmits  $X_{t}$ to the receiver, and $a_{t}=0$ otherwise. The
transmission decision is an instantaneous, error-free, single-bit
feedback from the receiver to the transmitter. We assume that   each
data transmission is packetized into $T_{u}$ packets and the transmission
of each packet takes one time slot.  The duration of a time step is
therefore not uniform.  Let $L(a_{t})$ denote the number of time
slots in time step $t$ with action $a_{t}$ being taken, which can
be expressed as:
\begin{align}
L(a_{t})=\begin{cases}
1, & \text{ if }a_{t}=0\\
T_{u}, & \text{ if }a_{t}=1
\end{cases} & .\label{Eq:time_step}
\end{align}

We assume that channel fading is constant within each time slot but
varies independently across different time slots. Additionally, channel
state information is assumed to be available only at the receiver,
and the transmitter transmits data packets at a constant rate. A memoryless
Bernoulli process $b_{t,i}\in\left\{ 0,1\right\} $ is used to characterize
transmission failures, where $b_{t,i}=1$ indicates a successful packet
transmission in the $i$th time slot of time step $t$, and $b_{t,i}=0$
indicates a transmission failure. The transmission success probability
of a packet is denoted by $p_{u}$. The receiver successfully receives
the data if all $T_{u}$ packets are successfully transmitted  within
the time step. Let $b_{t}\in\left\{ 0,1\right\} $ denote the transmission
status of the data at time step $t$, defined as  $b_{t}=\mathbf{\prod}_{i=1}^{T_{u}}b_{t,i}$,
where $b_{t}=1$ indicates that the transmission is successful, and
$b_{t}=0$ otherwise. Therefore, the transmission success probability
of the data is given by $\mathrm{Pr}(b_{t}=1)=p_{u}^{T_{u}}$, and
the transmission failure probability of the data is given by $\mathrm{Pr}(b_{t}=0)=1-p_{u}^{T_{u}}$.

When the data $X_{t}$ arrives at the receiver, the validator processes
it using a large binary classifier to determine its relevance to the
downstream task.  The validator provides the identification result
$H(X_{t})$, which is assumed to be perfect,  i.e., $H(X_{t})=Y_{t}$.
 We define $d_{t}\in\{0,1\}$ as an indicator of successful monitoring
at time step $t$.  $d_{t}=1$ if $X_{t}$ is successfully received
and the identification result $H(X_{t})$ matches the target (i.e.,
$b_{t}=1$ and $H(X_{t})=1$); otherwise, $d_{t}=0$.  In particular,
the probability of successful   monitoring  is given by:
\begin{align}
 & \mathrm{Pr}(d_{t}=1)\nonumber \\
= & \mathrm{Pr}(H(X_{t})=1)\mathrm{Pr}(b_{t}=1)\nonumber \\
= & \mathrm{Pr}(Y_{t}=1)\mathrm{Pr}(b_{t}=1)\nonumber \\
= & \{(1-\hat{p}_{A})\mathrm{Pr}(F(X_{t})=1)+\hat{p}_{B}\mathrm{Pr}(F(X_{t})=0)\}\mathrm{Pr}(b_{t}=1),
\end{align}

\noindent where
\begin{align}
\hat{p}_{A} & \triangleq\mathrm{Pr}(Y_{t}=0|F(X_{t})=1)=\dfrac{(1-q)p_{A}}{(1-q)p_{A}+q(1-p_{B})},\label{Eq:misclassification_2}\\
\hat{p}_{B} & \triangleq\mathrm{Pr}(Y_{t}=1|F(X_{t})=0)=\dfrac{qp_{B}}{(1-q)(1-p_{A})+qp_{B}}.\label{Eq:misclassification_3}
\end{align}

AoI has been extensively used to quantify the freshness of data perceived
by the receiver, thereby enhancing the utility of decision-making
processes \cite{Yates2021}. These efforts are driven by the consensus
that fresher data generally provides  more valuable information. However,
AoI does not directly measure data content or its relevance to the
downstream task.  The TAoI  proposed in this work  differs from AoI
 by considering not only  the time lag of received information  but
also its relevance to the downstream task. Specifically,  TAoI decreases
only when the monitoring task is successful; otherwise, it increases. 

\begin{figure}[!t]
\centering \includegraphics[width=0.45\textwidth]{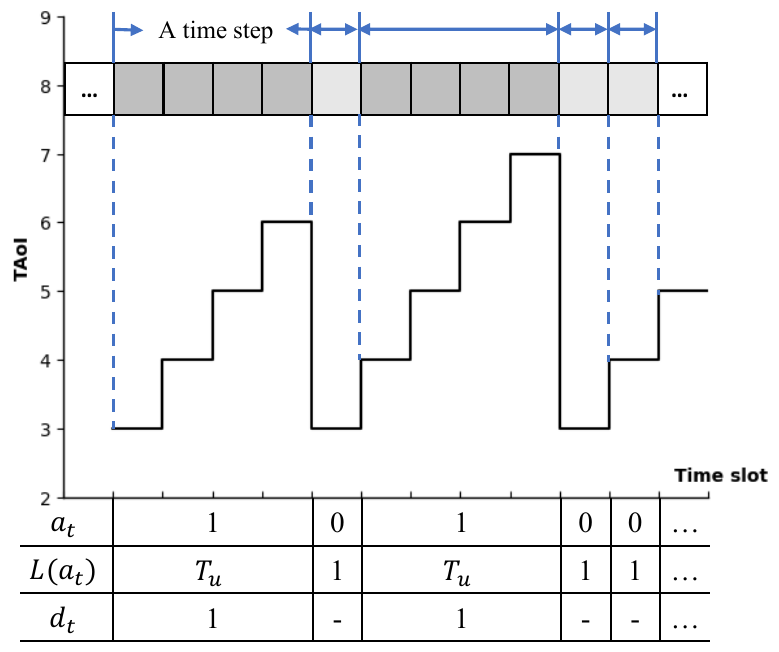}\caption{An illustration of the evolution of the TAoI ($T_{u}=3$).}
\label{Fig:TAoI_dynamic}
\end{figure}

Formally, let $U_{t}$ denote the time step at which the most recent
successfully monitored data was generated.  Then, the TAoI at the
$i$th time slot of time step $t$ is defined as:
\begin{align}
\Delta_{t,i}=\sum_{n=U_{t}}^{t-1}L(a_{n})+i-1 & ,\label{Eq:TAoI}
\end{align}

\noindent where the first term is the total number of time slots in
the previous time steps since $U_{t}$. For ease of exposition, we
represent the TAoI at the beginning of time step $t$ as $\Delta_{t}$.
That is, $\Delta_{t}=\Delta_{t,1}=\sum_{n=U_{t}}^{t-1}L(a_{n})$.
When the transmitter sends data to the receiver, TAoI is updated based
on the transmission outcome. If the received data is the target data
(i.e., $a_{t}=1$ and $d_{t}=1$),  TAoI decreases to $T_{u}$. If
the received data  is not the target data (i.e., $a_{t}=1$ and $d_{t}=0$),
 TAoI increases by $T_{u}$. If the decision maker does not  request
data transmission (i.e., $a_{t}=0$), TAoI increases by one. Thus,
the dynamics of TAoI can be expressed as follows:
\begin{align}
\Delta_{t+1}=\begin{cases}
T_{u}, & \text{ if }a_{t}=1\;\text{and}\;d_{t}=1;\\
\min\{\Delta_{t}+T_{u},\hat{\Delta}\}, & \text{ if }a_{t}=1\;\text{and}\;d_{t}=0;\\
\min\{\Delta_{t}+1,\hat{\Delta}\}, & \text{ if }a_{t}=0,
\end{cases}\label{Eq:AoI}
\end{align}
where $\hat{\Delta}$ is the finite upper limit of the TAoI.  An example
of the TAoI evolution with $T_{u}=3$ is illustrated in Fig. \ref{Fig:TAoI_dynamic}.

\section{SMDP Formulation and Solution}

\label{Sec:Section 4}

\subsection{SMDP Formulation}

The dynamic transmission problem in this work involves non-constant
time intervals between decision instants, making SMDP a suitable framework.
 We therefore formulate this problem as an infinite time-horizon SMDP,
defined by the tuple $\left(\mathcal{S},\mathcal{A},t^{+},\mathrm{Pr}(\cdot,\cdot),C(\cdot,\cdot)\right)$.
We depict each element as follows:

\subsubsection{State space $\mathcal{S}$}

The state $\mathbf{s}_{t}$ of the SMDP at time step $t$ is defined
as $\mathbf{s}_{t}\triangleq(\Delta_{t},F(X_{t}))$, where $\Delta_{t}$
denotes the TAoI at the beginning of time step $t$ and $F(X_{t})$
represents the pre-identification result for  $X_{t}$ at time step
$t$. The set of all possible states is denoted by $\mathcal{S}$.
It is finite because the TAoI is bounded by the upper limit $\hat{\Delta}$.

\subsubsection{Action space $\mathcal{A}$}

The action at time step $t$ is the transmission decision $a_{t}$
and the action space is $\mathcal{A}\triangleq\{0,1\}$.

\subsubsection{Decision epoch $t^{+}$}

Transmission decisions are made at the beginning of each time step.
As detailed in (\ref{Eq:time_step}), the time interval $L(a_{t})$
between two adjacent decision epochs depends on the action $a_{t}$
taken at time step $t$.

\subsubsection{Transition probability $\mathrm{Pr}(\cdot,\cdot)$}

Given the current state $\mathbf{s}_{t}=(\Delta_{t},F(X_{t}))$ and
action $a_{t}$, the transition probability to the next state $\mathbf{s}_{t+1}=(\Delta_{t+1},F(X_{t+1}))$
is denoted by $\mathrm{Pr}(\mathbf{s}_{t+1}|\mathbf{s}_{t},a_{t})$.
The transition probabilities are defined according to the TAoI evolution
dynamic in (\ref{Eq:AoI}) and are further detailed in Table \ref{Ta:Tran_pro-1}.

\subsubsection{Cost function $C(\cdot,\cdot)$}

The instantaneous cost $C(\mathbf{s}_{t},a_{t})$  under state $\mathbf{s}_{t}$
and action $a_{t}$ is defined as:
\begin{align}
C(\mathbf{s}_{t},a_{t}) & =C((\Delta_{t},F(X_{t})),a_{t})\nonumber \\
 & =\sum_{i=1}^{L(a_{t})}\Delta_{t,i}=\sum_{i=1}^{L(a_{t})}\Delta_{t}+i-1.
\end{align}

\begin{table}[!t]
\global\long\def\arraystretch{1.5}%
\centering \caption{Transition probability}
\begin{tabular}{lccc}
\toprule 
$\mathrm{Pr}(\mathbf{s}_{t+1}|\mathbf{s}_{t},a_{t})$ & $\mathbf{s}_{t}$ & $a_{t}$ & $\mathbf{s}_{t+1}$\tabularnewline
\midrule 
$(1-\hat{p}_{A})p_{u}^{T_{u}}g$ & $(\Delta_{t},1)$ & 1 & $(T_{u},1)$\tabularnewline
$\hat{p}_{B}p_{u}^{T_{u}}g$ & $(\Delta_{t},0)$ & 1 & $(T_{u},1)$\tabularnewline
$(1-\hat{p}_{A})p_{u}^{T_{u}}(1-g)$ & $(\Delta_{t},1)$ & 1 & $(T_{u},0)$\tabularnewline
$\hat{p}_{B}p_{u}^{T_{u}}(1-g)$ & $(\Delta_{t},0)$ & 1 & $(T_{u},0)$\tabularnewline
$(1-p_{u}^{T_{u}}+\hat{p}_{A}p_{u}^{T_{u}})g$ & $(\Delta_{t},1)$ & 1 & $(\min\{\Delta_{t}+T_{u},\hat{\Delta}\},1)$\tabularnewline
$(1-\hat{p}_{B}p_{u}^{T_{u}})g$ & $(\Delta_{t},0)$ & 1 & $(\min\{\Delta_{t}+T_{u},\hat{\Delta}\},1)$\tabularnewline
$(1-p_{u}^{T_{u}}+\hat{p}_{A}p_{u}^{T_{u}})(1-g)$ & $(\Delta_{t},1)$ & 1 & $(\min\{\Delta_{t}+T_{u},\hat{\Delta}\},0)$\tabularnewline
$(1-\hat{p}_{B}p_{u}^{T_{u}})(1-g)$ & $(\Delta_{t},0)$ & 1 & $(\min\{\Delta_{t}+T_{u},\hat{\Delta}\},0)$\tabularnewline
$g$ & $(\Delta_{t},1)$ & 0 & $(\min\{\Delta_{t}+1,\hat{\Delta}\},1)$\tabularnewline
$g$ & $(\Delta_{t},0)$ & 0 & $(\min\{\Delta_{t}+1,\hat{\Delta}\},1)$\tabularnewline
$1-g$ & $(\Delta_{t},1)$ & 0 & $(\min\{\Delta_{t}+1,\hat{\Delta}\},0)$\tabularnewline
$1-g$ & $(\Delta_{t},0)$ & 0 & $(\min\{\Delta_{t}+1,\hat{\Delta}\},0)$\tabularnewline
\bottomrule
\end{tabular}\label{Ta:Tran_pro-1}
\end{table}

In this paper, we aim to determine an optimal transmission policy,
denoted as $\pi=\{a_{1},a_{2},\cdots\}$, that minimizes the long-term
average TAoI. Given an initial system state $\mathbf{s}_{1}$, the
dynamic transmission problem can be formulated as:
\begin{align}
\min_{\pi}\limsup_{T\rightarrow\infty}\dfrac{\mathbb{E}\left[\sum_{t=1}^{T}C(\mathbf{s}_{t},a_{t})\mid\mathbf{s}_{1}\right]}{\mathbb{E}\left[\sum_{t=1}^{T}L(a_{t})\right]} & .\label{Eq:trans_policy}
\end{align}
Due to the non-uniform duration of time steps, the average cost in
(\ref{Eq:trans_policy}) is defined as the limit of the expected total
cost over a finite number of time steps, normalized by the expected
cumulative duration of these time steps.  

To address this problem, we first employ uniformization technique
to transform the SMDP into an equivalent discrete-time MDP \cite{Puterman2014,Tijms2003}.
The state and action spaces of the transformed MDP, denoted $\hat{\mathcal{S}}$
and $\hat{\mathcal{A}}$, respectively, remain identical to those
of the original SMDP, i.e., $\hat{\mathcal{S}}=\mathcal{S}$ and $\hat{\mathcal{A}}=\mathcal{A}$.
 For any $\mathbf{s}=(\Delta,F(X))\in\hat{\mathcal{S}}$ and $a\in\hat{\mathcal{A}}$,
the cost in the MDP is defined as:
\begin{align}
\bar{C}((\Delta,F(X)),a)=\Delta+\dfrac{1}{2}(L(a)-1) & ,\label{Eq:reward}
\end{align}
and the transition probability is given by:
\begin{align}
\bar{p}(\mathbf{s}^{\prime}|\mathbf{s},a)=\begin{cases}
\frac{\epsilon}{L(a)}p(\mathbf{s}^{\prime}|\mathbf{s},a), & \mathbf{s}^{\prime}\neq\mathbf{s}\\
1-\frac{\epsilon}{L(a)}, & \mathbf{s}^{\prime}=\mathbf{s}
\end{cases} & ,\label{Eq:Pro}
\end{align}
where $\epsilon$ is selected such that $\ensuremath{0<\epsilon\leq\min_{a}L(a)}$. 

 The transformed MDP is a finite-state, finite-action, average-cost
MDP. We first verify the existence of a deterministic stationary optimal
policy.  Per \cite[Theorem 8.4.5]{Puterman2014}, such a policy exists
for a finite-state, finite-action, average-cost MDP if the cost function
is bounded and the MDP is unichain. To confirm this, we assess two
conditions. Firstly, the cost function $\bar{C}((\Delta,F(X)),a)$
is bounded, as the TAoI is capped by an upper bound $\hat{\Delta}$.
Secondly, since the state $(\hat{\Delta},F(X))$, where $F(X)\in\{0,1\}$,
is reachable from all other states, and the states $(\hat{\Delta},0)$
and $(\hat{\Delta},1)$ mutually reachable via intermediate states.
This ensures the induced Markov chain has a single recurrent class,
establishing that the MDP is unichain. Thus, a deterministic stationary
optimal policy exists.

The average cost optimal policy $\pi^{*}$ can be obtained by solving
the average Bellman optimality equation, as given in \cite{Bertsekas2012}:
\begin{align}
V^{*}+h(\mathbf{s})=\min_{a\in\mathcal{A}}\left\{ \bar{C}(\mathbf{s},a)+\sum_{\mathbf{s}^{\prime}\in\mathcal{S}}\bar{p}(\mathbf{s}^{\prime}|\mathbf{s},a)h(\mathbf{s}^{\prime})\right\} ,\;\forall\mathbf{s}\in\mathcal{S} & ,\label{Eq:Bellman}
\end{align}
where $V^{*}$ represents the optimal average cost of (\ref{Eq:trans_policy})
for all initial states, and $h(\mathbf{s})$ is the relative value
function of the  MDP. We define the state-action value function as:
\begin{align}
Q(\mathbf{s},a)=\bar{C}(\mathbf{s},a)+\sum_{\mathbf{s}^{\prime}\in\mathcal{S}}\bar{p}(\mathbf{s}^{\prime}|\mathbf{s},a)h(\mathbf{s}^{\prime}),\;\forall\mathbf{s}\in\mathcal{S},a & \in\mathcal{A}\text{.}\label{Eq:Bellman-1}
\end{align}
Then, the optimal policy $\pi^{*}$ for any $\mathbf{s}\in\mathcal{S}$
can be given by:
\begin{equation}
\pi^{*}(\mathbf{s})=\arg\min_{a\in\mathcal{A}}Q(\mathbf{s},a),\;\forall\mathbf{s}\in\mathcal{S}.\label{Eq:opt_polic}
\end{equation}

\subsection{Structural Analysis and Algorithm Design}

 The optimal policy $\pi^{*}$ can be computed using the Relative
Value Iteration (RVI) algorithm.  Let $Q_{k}(\mathbf{s},a)$ and $h_{k}(\mathbf{s})$
denote the state-action value function and the relative value function
at the $k$-th iteration, respectively. The RVI  updates at the $k$-th
iteration  are defined as follows:
\begin{align}
Q_{k}(\mathbf{s},a)=\bar{C}(\mathbf{s},a)+\sum_{\mathbf{s}^{\prime}\in\mathcal{S}}\bar{p}(\mathbf{s}^{\prime}|\mathbf{s},a)h_{k-1}(\mathbf{s}^{\prime}) & ,\label{Eq:RVI-1-1}
\end{align}
and 
\begin{align}
h_{k}(\mathbf{s})=V_{k}(\mathbf{s})-V_{k}(\mathbf{s^{\dagger}}),\label{Eq:RVI-3-1}
\end{align}
where $V_{k}(\mathbf{s})=\min_{a\in\mathcal{A}}Q_{k}(\mathbf{s},a)$
is the state value function at the $k$-th iteration, and $\mathbf{s^{\dagger}}$
is a fixed reference state. The RVI algorithm converges to $Q(\mathbf{s},a)$
and $h(\mathbf{s})$ as $t\rightarrow\infty$, regardless of initialization.
However, achieving an exact solution requires infinitely many iterations,
rendering it impractical. To address this, we exploit the system\textquoteright s
structural properties to characterize the optimal policy efficiently.
Before presenting the theorem on these properties, we establish key
attributes of the relative value function $h(\mathbf{s})$ through
the following lemmas. 
\begin{lem}
\label{LM:Lemma1}For any given $F(X)$, the relative value function
$h(\Delta,F(X))$ is non-decreasing in $\Delta$.
\end{lem}
\begin{IEEEproof}
\label{PR:proof_lemma1} We prove Lemma 1 based on the RVI algorithm.
RVI converges as $k\rightarrow\infty$,  yielding $V_{k}(\mathbf{s})\rightarrow V(\mathbf{s})$
and $h_{k}(\mathbf{s})\rightarrow h(\mathbf{s})$. In the limit, 
(\ref{Eq:RVI-3-1}) simplifies to $h(\mathbf{s})=V(\mathbf{s})-V(\mathbf{s^{\dagger}})$.
Since $V(\mathbf{s^{\dagger}})$ is independent of $s$, $V(\mathbf{s})$
and $h(\mathbf{s})$ share the same monotonicity properties. Therefore,
to prove the lemma, it suffices to demonstrate the monotonicity of
$V_{k}(\mathbf{s})$ with respect to $\Delta$ for all $k\geq0$.
Specifically, we will show that for any two states $\mathbf{s}_{1}=(\Delta_{1},F(X))$
and $\mathbf{s}_{2}=(\Delta_{2},F(X))$ in $\mathcal{S}$, if $\Delta_{1}\leq\Delta_{2}$,
then $V_{k}(\mathbf{s}_{1})\leq V_{k}(\mathbf{s}_{2}),\;k=0,1,\cdots.$

We prove  the monotonicity of $V_{k}(\mathbf{s})$ with respect to
$\Delta$ by using mathematical induction. Without loss of generality,
we set $V_{0}(\mathbf{s})=0$ for all $s\in\mathcal{S}$, ensuring
that the monotonicity of $V_{k}(\mathbf{s})$ is satisfied at $k=0$.
Then, assuming that the monotonicity of $V_{k}(\mathbf{s})$ holds
up to $k>0$, we verify whether it holds for $k+1$. 

When $a=0$, it follows that
\begin{align}
 & Q_{k+1}(\mathbf{s}_{2},0)-Q_{k+1}(\mathbf{s}_{1},0)\nonumber \\
= & \Delta_{2}-\Delta_{1}+(1-\epsilon)(h_{k}(\mathbf{s}_{2})-h_{k}(\mathbf{s}_{1}))\nonumber \\
 & +\epsilon(1-g)(h_{k}(\Delta_{2}+1,0)-h_{k}(\Delta_{1}+1,0))\nonumber \\
 & +\epsilon g(h_{k}(\Delta_{2}+1,1)-h_{k}(\Delta_{1}+1,1))\overset{(a)}{\geq}0,\label{Eq:mono1}
\end{align}
where (a) follows the non-decreasing property of $h_{k}(\mathbf{s})$.
Thus, it can be easily deduced that $Q_{k+1}(\mathbf{s}_{1},0)\leq Q_{k+1}(\mathbf{s}_{2},0)$.

When $a=1$, it follows that
\begin{align}
 & Q_{k+1}(\mathbf{s}_{2},1)-Q_{k+1}(\mathbf{s}_{1},1)\nonumber \\
= & \Delta_{2}-\Delta_{1}+\left(1-\frac{\epsilon}{T_{u}}\right)(h_{k}(\mathbf{s}_{2})-h_{k}(\mathbf{s}_{1}))\nonumber \\
 & +\frac{\epsilon}{T_{u}}p_{1}(1-g)(h_{k}(\Delta_{2}+T_{u},0)-h_{k}(\Delta_{1}+T_{u},0))\nonumber \\
 & +\frac{\epsilon}{T_{u}}p_{1}g(h_{k}(\Delta_{2}+T_{u},1)-h_{k}(\Delta_{1}+T_{u},1))\geq0,\label{Eq:mono2}
\end{align}
where $p_{1}=1-\hat{p}_{B}p_{u}^{T_{u}}$ if $F(X)=0$,  and $p_{1}=1-p_{u}^{T_{u}}+\hat{p}_{A}p_{u}^{T_{u}}$
if $F(X)=1$. Similarly, we can obtain that $Q_{k+1}(\mathbf{s}_{1},1)\leq Q_{k+1}(\mathbf{s}_{2},1)$
due to the non-decreasing property of $h_{k}$. 

Finally, since $V_{k+1}(\mathbf{s})=\min_{a\in\mathcal{A}}Q_{k+1}(\mathbf{s},a)$,
we can obtain that $V_{k+1}(\mathbf{s}_{1})\leq V_{k+1}(\mathbf{s}_{2})$
for any $k$. Then, for any $k$, we also have $h_{k+1}(\mathbf{s}_{1})\leq h_{k+1}(\mathbf{s}_{2})$.
This concludes the proof of Lemma \ref{LM:Lemma1}.
\end{IEEEproof}
\begin{lem}
\label{LM:Lemma2} For any fixed $F(X)$, the relative value function
$h(\Delta,F(X))$ is concave in $\Delta$.
\end{lem}
\begin{IEEEproof}
\label{PR:proof_lemma2} The concavity of $h(\mathbf{s})$ with respect
to $\Delta$ for any given $F(X)$ can be demonstrated by showing
that, for any $\mathbf{s}_{1}=(\Delta_{1},F(X))$, $\mathbf{s}_{2}=(\Delta_{2},F(X))$,
$\mathbf{s}_{1}^{\prime}=(\Delta_{1}+w,F(X))$, and $\mathbf{s}_{2}^{\prime}=(\Delta_{2}+w,F(X))$
in $\mathcal{S}$, if $\Delta_{1}\leq\Delta_{2}$, then
\begin{align}
 & h_{k}(\Delta_{1}+w,F(X))-h_{k}(\Delta_{1},F(X))\geq\label{Eq:concave}\\
 & h_{k}(\Delta_{2}+w,F(X))-h_{k}(\Delta_{2},F(X)),k=0,1,\dots,\nonumber 
\end{align}
where $w$ is a positive integer.  Without loss of generality, we
set $V_{0}(\mathbf{s})=0$ and $h_{0}(\mathbf{s})=0$ for all $\mathbf{s}\in\mathcal{S}$,
ensuring that (\ref{Eq:concave}) holds at $k=0$. Then, we assume
that (\ref{Eq:concave}) holds for all $k>0$ and investigate whether
it holds for $k+1$. 

 To determine the concavity of $h_{k+1}(\mathbf{s})$, we can verify
the concavity of $Q_{k+1}(\mathbf{s},a)$ with respect to $\Delta$.
For convenience, we introduce $\bar{Q}(\mathbf{s}^{\prime},\mathbf{s},a)=Q(\mathbf{s}^{\prime},a)-Q(\mathbf{s},a)$,
where $\mathbf{s}^{\prime}=(\Delta+w,F(X))$.

When $a=0$, it follows that 
\begin{align}
 & \bar{Q}_{k+1}(\mathbf{s}_{1}^{\prime},\mathbf{s}_{1},0)-\bar{Q}_{k+1}(\mathbf{s}_{2}^{\prime},\mathbf{s}_{2},0)\nonumber \\
= & (1-\epsilon)[(h_{k}(\Delta_{1}+w,F(X))-h_{k}(\Delta_{1},F(X)))\nonumber \\
 & -(h_{k}(\Delta_{2}+w,F(X))-h_{k}(\Delta_{2},F(X)))]\nonumber \\
 & +\epsilon(1-g)[(h_{k}(\Delta_{1}+w+1,0)-h_{k}(\Delta_{1}+1,0))\nonumber \\
 & -(h_{k}(\Delta_{2}+w+1,0)-h_{k}(\Delta_{2}+1,0))]\nonumber \\
 & +\epsilon g[(h_{k}(\Delta_{1}+w+1,1)-h_{k}(\Delta_{1}+1,1))\nonumber \\
 & -(h_{k}(\Delta_{2}+w+1,1)-h_{k}(\Delta_{2}+1,1))]\overset{(a)}{\geq}0,
\end{align}
where (a) follows the concavity of $h_{k}(\mathbf{s})$.  Thus, $Q_{k+1}(\mathbf{s},0)$
is concave in $\Delta$ for any given $F(X)$.

When $a=1$, it follows that 
\begin{align}
 & \bar{Q}_{k+1}(\mathbf{s}_{1}^{\prime},\mathbf{s}_{1},1)-\bar{Q}_{k+1}(\mathbf{s}_{2}^{\prime},\mathbf{s}_{2},1)\nonumber \\
= & (1-\frac{\epsilon}{T_{u}})[(h_{k}(\Delta_{1}+w,F(X))-h_{k}(\Delta_{1},F(X)))\nonumber \\
 & -(h_{k}(\Delta_{2}+w,F(X))-h_{k}(\Delta_{2},F(X)))]\nonumber \\
 & +\frac{\epsilon p_{1}}{T_{u}}(1-g)[(h_{k}(\Delta_{1}+w+T_{u},0)-h_{k}(\Delta_{1}+T_{u},0))\nonumber \\
 & -(h_{k}(\Delta_{2}+w+T_{u},0)-h_{k}(\Delta_{2}+T_{u},0)]\nonumber \\
 & +\frac{\epsilon p_{1}}{T_{u}}g[(h_{k}(\Delta_{1}+w+T_{u},1)-h_{k}(\Delta_{1}+T_{u},1))\nonumber \\
 & -(h_{k}(\Delta_{2}+w+T_{u},1)-h_{k}(\Delta_{2}+T_{u},1)]\geq0.
\end{align}
Thus, $Q_{k+1}(\mathbf{s},1)$ is concave in $\Delta$ for any given
$F(X)$.

Since both $Q_{k+1}(\mathbf{s},0)$ and $Q_{k+1}(\mathbf{s},1)$ are
 concave functions with respect to $\Delta$, $V_{k+1}(\mathbf{s})=\min_{a\in\mathcal{A}}Q_{k+1}(\mathbf{s},a)$
is a concave functions in $\Delta$, and so does $h_{k+1}(\mathbf{s})$
 for any given $F(X)$.  This concludes the proof of Lemma \ref{LM:Lemma2}.
\end{IEEEproof}
By Lemmas 1 and 2, the slope of $h(\Delta,F(X))$ with respect to
$\Delta$ is non-negative and non-increasing. Consequently, there
exists a lower bound on this slope, which is explicitly characterized
as follows.
\begin{lem}
\label{LM:Lemma3} For for any fixed $F(X)$ and any states $\mathbf{s}_{1}=(\Delta_{1},F(X))$
and $\mathbf{s}_{2}=(\Delta_{2},F(X))$ in $\mathcal{S}$, such that
$\Delta_{1}\leq\Delta_{2}$, the relative value function $h(\mathbf{s})$
satisfies: 
\begin{equation}
h(\Delta_{2},F(X))-h(\Delta_{1},F(X))\geq\dfrac{L(a)}{\epsilon(1-p_{1})}(\Delta_{2}-\Delta_{1}),
\end{equation}
where $p_{1}=1-p_{u}^{T_{u}}+\hat{p}_{A}p_{u}^{T_{u}}$ if $F(X)=1$,
and $p_{1}=1-\hat{p}_{B}p_{u}^{T_{u}}$ if $F(X)=0$.
\end{lem}
\begin{IEEEproof}
\label{PR:proof_lemma3} We will establish this property by induction
on $k$. The lower bound of $h(\mathbf{s}_{2})-h(\mathbf{s}_{1})$
can be proven by showing that for any fixed $F(X)$ and any state
$\mathbf{s}_{1}=(\Delta_{1},F(X))$, $\mathbf{s}_{2}=(\Delta_{2},F(X))\in\mathcal{S}$,
such that $\Delta_{1}\leq\Delta_{2}$, 
\begin{align}
 & h_{k}(\Delta_{2},F(X))-h_{k}(\Delta_{1},F(X))\nonumber \\
 & \geq\frac{L(a)}{\epsilon(1-p_{1})}(\Delta_{2}-\Delta_{1}),\quad k=0,1,\dots.\label{eq:lower_bound}
\end{align}

First, we let $h_{0}(\mathbf{s})=\frac{L(a)}{\epsilon(1-p_{1})}\Delta$
for all $\mathbf{s}=(\Delta,F(X))\in\mathcal{S}$. This ensures that
(\ref{eq:lower_bound}) is satisfied for $k=0$. Assume that (\ref{eq:lower_bound})
holds up till $k>0$,  we then verify whether it holds for $k+1$.
Since $h_{k+1}(\mathbf{s}_{2})-h_{k+1}(\mathbf{s}_{1})=V_{k+1}(\mathbf{s}_{2})-V_{k+1}(\mathbf{s}_{1})$
 and $V_{k+1}(\mathbf{s})=\min_{a\in\mathcal{A}}Q_{k+1}(\mathbf{s},a)$,
we investigate the state-action value functions in the following.
We first consider the case that $F(X)=1$. 

When $a=0$, we have
\begin{align}
 & \bar{Q}_{k+1}(\mathbf{s}_{2},\mathbf{s}_{1},0)\nonumber \\
= & Q_{k+1}((\Delta_{2},1),0)-Q_{k+1}((\Delta_{1},1),0)\nonumber \\
= & \Delta_{2}-\Delta_{1}+\left(1-\epsilon\right)(h_{k}(\Delta_{2},1)-h_{k}(\Delta_{1},1))\nonumber \\
 & +\epsilon(1-g)(h_{k}(\Delta_{2}+1,0)-h_{k}(\Delta_{1}+1,0))\nonumber \\
 & +\epsilon g(h_{k}(\Delta_{2}+1,1)-h_{k}(\Delta_{1}+1,1))\nonumber \\
\geq & (\Delta_{2}-\Delta_{1})+\frac{L(a)}{\epsilon(1-p_{1})}(\Delta_{2}-\Delta_{1})\nonumber \\
= & \left(1+\frac{L(a)}{\epsilon(1-p_{1})}\right)(\Delta_{2}-\Delta_{1})\nonumber \\
\geq & \frac{L(a)}{\epsilon(1-p_{1})}(\Delta_{2}-\Delta_{1}).
\end{align}

When  $a=1$, we have
\begin{align}
 & \bar{Q}_{k+1}(\mathbf{s}_{2},\mathbf{s}_{1},1)\nonumber \\
= & Q_{k+1}((\Delta_{2},1),1)-Q_{k+1}((\Delta_{1},1),1)\nonumber \\
= & \Delta_{2}-\Delta_{1}+\left(1-\frac{\epsilon}{T_{u}}\right)h_{k}(\Delta_{2},1)-h_{k}(\Delta_{1},1)\nonumber \\
 & +\frac{\epsilon}{T_{u}}p_{1}(1-g)\left(h_{k}(\Delta_{2}+T_{u},0)-h_{k}(\Delta_{1}+T_{u},0)\right)\nonumber \\
 & +\frac{\epsilon}{T_{u}}p_{1}g\left(h_{k}(\Delta_{2}+T_{u},1)-h_{k}(\Delta_{1}+T_{u},1)\right)\nonumber \\
\geq & \Delta_{2}-\Delta_{1}+\left(1-\frac{\epsilon}{T_{u}}+\frac{\epsilon}{T_{u}}p_{1}\right)\frac{L(a)}{\epsilon(1-p_{1})}(\Delta_{2}-\Delta_{1})\nonumber \\
\overset{(a)}{=} & \frac{L(a)}{\epsilon(1-p_{1})}(\Delta_{2}-\Delta_{1}),
\end{align}
where (a) follows from $L(a)=T_{u}$ when $a=1$ according to (\ref{Eq:time_step}).
Therefore, we can prove that (\ref{eq:lower_bound}) holds for any
$k$ when the optimal actions in $\mathbf{s}_{1}$ and $\mathbf{s}_{2}$
are the same.

When the optimal policies in $\mathbf{s}_{1}$ and $\mathbf{s}_{2}$
are different, i.e., $a_{1}$ and $a_{2}$, we have
\begin{align}
 & h_{k+1}(\Delta_{2},F(X))-h_{k+1}(\Delta_{1},F(X))\nonumber \\
= & Q_{k+1}((\Delta_{2},F(X)),a_{2})-Q_{k+1}((\Delta_{1},F(X)),a_{1})\nonumber \\
\geq & Q_{k+1}((\Delta_{2},F(X)),a_{2})-Q_{k+1}((\Delta_{1},F(X)),a_{2})\nonumber \\
\geq & \frac{L(a)}{\epsilon(1-p_{1})}(\Delta_{2}-\Delta_{1}).
\end{align}
It shows that (\ref{eq:lower_bound}) also holds for any $k$ in this
case.

Altogether, we can conclude that (\ref{eq:lower_bound}) holds for
any $k$ with $F(X)=1$. By following the same analysis, we can prove
the same result with $F(X)=0$. By induction, we have $h(\mathbf{s}_{2})-h(\mathbf{s}_{1})\geq\frac{L(a)}{\epsilon(1-p_{1})}(\Delta_{2}-\Delta_{1})$
for any fixed $F(X)$. This concludes the proof of Lemma \ref{LM:Lemma3}.
\end{IEEEproof}
Based on Lemmas \ref{LM:Lemma1}-\ref{LM:Lemma3}, we  derive the
structure of the optimal transmission policy  in the following theorem.
\begin{thm}
\label{Th:Theorem1} For any fixed $F(X)$, there exists a deterministic
stationary optimal policy that is of threshold-type in $\Delta$.
Specifically, if $\pi^{*}(\Delta_{1},F(X))=1$, then $\pi^{*}(\Delta_{2},F(X))=1$
for all $\Delta_{1}\leq\Delta_{2}$.
\end{thm}
\begin{IEEEproof}
\label{PR:proof_Theorem1} For any $\mathbf{s}_{1}=(\Delta_{1},1)$
and $\mathbf{s}_{2}=(\Delta_{2},1)$ in $\mathcal{S}$ with $\Delta_{1}\leq\Delta_{2}$,
we have
\begin{align}
 & Q((\Delta_{2},1),1)-Q((\Delta_{1},1),1)\nonumber \\
= & \Delta_{2}-\Delta_{1}+\left(1-\frac{\epsilon}{L(a)}\right)\left(h(\Delta_{2},1)-h(\Delta_{1},1)\right)\nonumber \\
 & +\frac{\epsilon p_{1}}{L(a)}(1-g)\left(h(\Delta_{2}+L(a),0)-h(\Delta_{1}+L(a),0)\right)\nonumber \\
 & +\frac{\epsilon p_{1}}{L(a)}g\left(h(\Delta_{2}+L(a),1)-h(\Delta_{1}+L(a),1)\right)\nonumber \\
\overset{(a)}{\leq} & \Delta_{2}-\Delta_{1}+\left(1-\frac{\epsilon}{L(a)}\right)\left(h(\Delta_{2},1)-h(\Delta_{1},1)\right)\nonumber \\
 & +\frac{\epsilon p_{1}}{L(a)}(1-g)\left(h(\Delta_{2},0)-h(\Delta_{1},0)\right)\nonumber \\
 & +\frac{\epsilon p_{1}}{L(a)}g\left(h(\Delta_{2},1)-h(\Delta_{1},1)\right)\nonumber \\
= & \Delta_{2}-\Delta_{1}+h(\Delta_{2},1)-h(\Delta_{1},1)\nonumber \\
 & -\frac{\epsilon p_{1}}{L(a)}(g-1)\left(h(\Delta_{2},0)-h(\Delta_{1},0)\right)\nonumber \\
 & -\frac{\epsilon}{L(a)}(1-p_{1}g)\left(h(\Delta_{2},1)-h(\Delta_{1},1)\right)\nonumber \\
\overset{(b)}{\leq} & \left(1-\frac{p_{1}(g-1)}{1-p_{1}}-\frac{1-p_{1}g}{1-p_{1}}\right)(\Delta_{2}-\Delta_{1})\nonumber \\
 & +h(\Delta_{2},1)-h(\Delta_{1},1)\nonumber \\
= & h(\Delta_{2},1)-h(\Delta_{1},1)\label{Eq:theorem1}
\end{align}
where $(a)$ follows from the concavity of $h(\mathbf{s})$, and $(b)$
follows from the Lemma \ref{LM:Lemma3}. 

Eq. (\ref{Eq:theorem1}) implies that $V(\mathbf{s}_{2})-V(\mathbf{s}_{1})=h(\mathbf{s}_{2})-h(\mathbf{s}_{1})\geq Q(\mathbf{s}_{2},1)-Q(\mathbf{s}_{1},1)$.
 Suppose $\pi^{*}(\Delta_{1},1)=1$, then we obtain $V(\mathbf{s}_{1})=Q(\mathbf{s}_{1},1)$.
Thus, we have $V(\mathbf{s}_{2})\geq Q(\mathbf{s}_{2},1)$. However,
 the state value function definition, $V(\mathbf{s})=\min_{a\in\mathcal{A}}Q(\mathbf{s},a)$,
simultaneously requires $V(\mathbf{s}_{2})\leq Q(\mathbf{s}_{2},1)$,
forcing equality  $V(\mathbf{s}_{2})=Q(\mathbf{s}_{2},1)$ and consequently
$\pi^{*}(\Delta_{2},1)=1$.

Similarly, for any states $\mathbf{s}_{1}=(\Delta_{1},0)$, $\mathbf{s}_{2}=(\Delta_{2},0)$
in $\mathcal{S}$ with $\Delta_{1}\leq\Delta_{2}$, we can also proof
that if $\pi^{*}(\Delta_{1},0)=1$, then $\pi^{*}(\Delta_{2},0)=1$.
This concludes the proof of Theorem \ref{Th:Theorem1}.
\end{IEEEproof}
According to Theorem \ref{Th:Theorem1}, the optimal policy is characterized
 as a threshold policy, with $\Omega_{0}^{*}$ and $\Omega_{1}^{*}$
serving as the thresholds for $F(X)=0$ and $F(X)=1$, respectively.
Specifically, the policy is defined such that $\pi^{*}(\Delta,0)=1$
if $\Delta\geq\Omega_{0}^{*}$, and 0 otherwise; similarly, $\pi^{*}(\Delta,1)=1$
if $\Delta\geq\Omega_{1}^{*}$, and 0 otherwise. The relationship
between the thresholds $\Omega_{0}^{*}$ and $\Omega_{1}^{*}$ of
the optimal policy is given by Theorem \ref{Th:Theorem2-1}. Before
formally presenting the Corollary, we provide a necessary lemma.
\begin{figure}[!t]
\centering \includegraphics[width=0.49\textwidth]{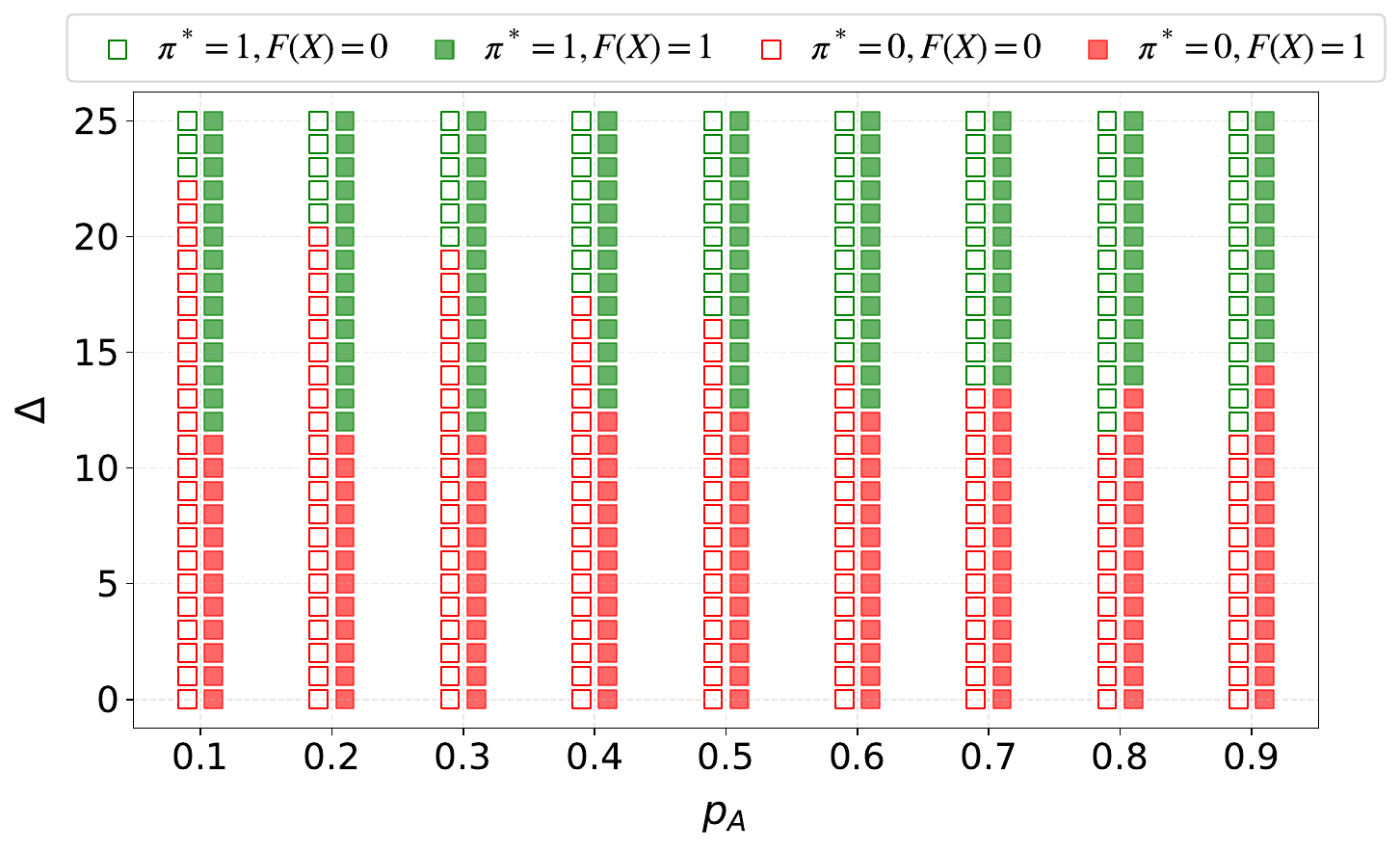}\caption{Structure of the optimal policy for different $p_{A}$ ($T_{u}=10$,
$q=0.7$, $p_{u}=1$, $p_{B}=0.3$).}
\label{Fig:optimal-thresholds-2D-pA}
\end{figure}

\begin{lem}
\label{LM:Lemma4}For any given $\Delta$, the relative value function
$h(\Delta,F(X))$ is decreasing in $F(X)$ if $1-p_{A}>p_{B}$, and
increasing in $F(X)$ if $1-p_{A}<p_{B}$.
\end{lem}
\begin{IEEEproof}
\label{PR:proof_lemma4} Similar to the proof of Lemma 1, the proof
of Lemma 4 only requires demonstrating the monotonicity of $V_{k}(\mathbf{s})$
with respect to $F(X)$ for all $k\geq0$. Specifically, for any two
states $\mathbf{s}_{0}=(\Delta,0)$ and $\mathbf{s}_{1}=(\Delta,1)$
in $\mathcal{S}$, if $1-p_{A}>p_{B}$, then $V_{k}(\mathbf{s}_{0})>V_{k}(\mathbf{s}_{1})$,
and if $1-p_{A}<p_{B}$, then $V_{k}(\mathbf{s}_{0})<V_{k}(\mathbf{s}_{1})$
for all $k\geq0$.

We use mathematical induction to prove the monotonicity of $V_{k}(\mathbf{s})$
with respect to $F(X)$. Without loss of generality, we set $V_{0}(\mathbf{s})=0$
for all $s\in\mathcal{S}$, ensuring that the monotonicity of $V_{k}(\mathbf{s})$
is satisfied at $k=0$. Then, assuming that the monotonicity of $V_{k}(\mathbf{s})$
holds up to $k>0$, we verify whether it holds for $k+1$. 

We first analyze the decreasing property of $V_{k}(\mathbf{s})$ with
respect to $F(X)$ when $1-p_{A}>p_{B}$.

When $a=0$, it follows that
\begin{align}
 & Q_{k+1}((\Delta,1),0)-Q_{k+1}((\Delta,0),0)\nonumber \\
= & (1-\epsilon)(h_{k}(\Delta,1)-h_{k}(\Delta,0))\overset{(a)}{<}0,
\end{align}
where $(a)$ follows the decreasing property of $h_{k}(\mathbf{s})$
in $F(X)$. Thus, it can be easily deduced that $Q_{k+1}(\mathbf{s}_{0},0)>Q_{k+1}(\mathbf{s}_{1},0)$.

When $a=1$, it follows that
\begin{align}
 & Q_{k+1}((\Delta,1),1)-Q_{k+1}((\Delta,0),1)\nonumber \\
= & \frac{\epsilon}{T_{u}}(p_{11}-p_{10})g\left[h_{k}(\Delta+T_{u},1)-h_{k}(T_{u},1)\right]\nonumber \\
 & +\frac{\epsilon}{T_{u}}(p_{11}-p_{10})(1-g)\left[h_{k}(\Delta+T_{u},0)-h_{k}(T_{u},0)\right]\nonumber \\
 & +\left(1-\frac{\epsilon}{T_{u}}\right)(h_{k}(\Delta,1)-h_{k}(\Delta,0))\overset{(b)}{<}0,
\end{align}
where $p_{10}=1-\hat{p}_{B}p_{u}^{T_{u}}$ , and $p_{11}=1-p_{u}^{T_{u}}+\hat{p}_{A}p_{u}^{T_{u}}$.
When $1-p_{A}>p_{B}$, we have $1-\hat{p}_{A}>\hat{p}_{B}$ according
to (\ref{Eq:misclassification_2}) and (\ref{Eq:misclassification_3}),
which implies $p_{10}>p_{11}$. Additionally, due to the non-decreasing
property of $h_{k}$ in $\Delta$ and the decreasing property of $h_{k}$
in $F(X)$ , we can conclude that $(b)$ holds, i.e., $Q_{k+1}(\mathbf{s}_{0},1)>Q_{k+1}(\mathbf{s}_{1},1)$
. 

Since $V_{k+1}(\mathbf{s})=\min_{a\in\mathcal{A}}Q_{k+1}(\mathbf{s},a)$,
we can obtain that $V_{k+1}(\mathbf{s}_{0})>V_{k+1}(\mathbf{s}_{1})$
for any $k$ when $1-p_{A}>p_{B}$. Then, for any $k$, we also have
$h_{k+1}(\mathbf{s}_{0})>h_{k+1}(\mathbf{s}_{1})$ when $1-p_{A}>p_{B}$.
Thus, $h(\Delta,F(X))$ is decreasing in $F(X)$ if $1-p_{A}>p_{B}$.

Similarly, $h(\Delta,F(X))$ is increasing in $F(X)$ if $1-p_{A}<p_{B}$.
This concludes the proof of Lemma \ref{LM:Lemma4}.
\end{IEEEproof}
\begin{thm}
For any given $\Delta$, the threshold $\Omega_{0}^{*}$ is larger
than the threshold $\Omega_{1}^{*}$ if $1-p_{A}>p_{B}$, the threshold
$\Omega_{0}^{*}$ is samller than the threshold $\Omega_{1}^{*}$
if $1-p_{A}<p_{B}$. In addition, when $1-p_{A}=p_{B}$, the thresholds
$\Omega_{0}^{*}$ and $\Omega_{1}^{*}$ of the optimal policy are
equal. That is, the optimal policy has a single optimal threshold,
which is independent of $F(X)$.\label{Th:Theorem2-1}
\end{thm}
\begin{IEEEproof}
\label{PR:proof_Theorem2-1}Let $\Omega_{F(X)}^{*}$ denote the threshold
associated with $F(X)$. That is, given $F(X)$, it is optimal to
transmit when $\Delta\geq\Omega_{F(X)}^{*}$. Let $\Delta_{0}=\Omega_{0}^{*}$,
we have
\begin{align}
 & \mbox{\small\ensuremath{Q((\Delta_{0},1),1)-Q((\Delta_{0},1),0)-\left[Q((\Delta_{0},0),1)-Q((\Delta_{0},0),0)\right]}}\nonumber \\
= & \left(\epsilon-\frac{\epsilon}{T_{u}}\right)\left[h(\Delta_{0},1)-h(\Delta_{0},0)\right]\nonumber \\
 & +\frac{\epsilon}{T_{u}}(p_{11}-p_{10})(1-g)\left[h_{k}(\Delta_{0}+T_{u},0)-h_{k}(T_{u},0)\right]\nonumber \\
 & +\frac{\epsilon}{T_{u}}(p_{11}-p_{10})g\left[h_{k}(\Delta_{0}+T_{u},1)-h_{k}(T_{u},1)\right].\label{eq:cor1}
\end{align}

When $1-p_{A}>p_{B}$, we have $p_{10}>p_{11}$ and the decreasing
property of $h_{k}$ in $F(X)$. Then, due to the non-decreasing property
of $h_{k}$ in $\Delta$, the above Eq. (\ref{eq:cor1}) is less than
$0$. Since $Q((\Delta_{0},0),1)\leq Q((\Delta_{0},0),0)$, we have
$Q((\Delta_{0},1),1)-Q((\Delta_{0},1),0)<0$. Therefore, we can deduce
that $\Delta_{0}=\Omega_{0}^{*}>\Omega_{1}^{*}$ if $1-p_{A}>p_{B}$.

Similarly, we obtain that $\Omega_{0}^{*}<\Omega_{1}^{*}$ if $1-p_{A}<p_{B}$.
Moreover, the value function is independent of $F(X)$ if $1-p_{A}=p_{B}$,
which means $h(\Delta_{0},1)=h(\Delta_{0},0)$. In this case, Eq.
(\ref{eq:cor1}) equals $0$, implying $\Omega_{0}^{*}=\Omega_{1}^{*}$.
This concludes the proof of Corollary \ref{Th:Theorem2-1}.
\end{IEEEproof}
\begin{algorithm}
\caption{Threshold-based Relative Value Iteration}
\label{alg:algorithm1-1} \begin{algorithmic}[1]

\STATE \textbf{Initialization:} Set $V(\mathbf{s})=0$, $h(\mathbf{s})=0$
and $V^{'}(\mathbf{s})=\infty$ for all $\mathbf{s}\in S$, select
a reference state $\mathbf{s^{\dagger}}$, and set $\bar{\lambda}$.

\REPEAT

\STATE For all $\mathbf{s}=\{\Delta,F(X)\}\in S$, $\Omega_{0}^{*}=\infty$,
$\Omega_{1}^{*}=\infty$;

\FOR {$\mathbf{s}\in S$}

\IF{$F(X)=0$ and $\text{\ensuremath{\Delta}}\geq\Omega_{0}^{*}$
}

\STATE $\pi^{*}=1$;

\ELSIF{$F(X)=1$ and $\text{\ensuremath{\Delta}}\geq\Omega_{1}^{*}$
}

\STATE $\pi^{*}=1$;

\ELSE

\STATE $\pi^{*}=\arg\min\limits _{a\in\mathcal{A}}\left\{ \bar{C}(\mathbf{s},\pi^{*})+\sum\limits _{\mathbf{s}^{\prime}\in\mathcal{S}}\bar{p}(\mathbf{s}^{\prime}|\mathbf{s},\pi^{*})h(\mathbf{s}^{\prime})\right\} $;

\IF{$F(X)=0$ and $\pi^{*}=1$ }

\STATE $\Omega_{0}^{*}=\Delta$;

\ELSIF{$F(X)=1$ and $\pi^{*}=1$ }

\STATE $\Omega_{1}^{*}=\Delta$;

\ENDIF

\ENDIF

\STATE $V(\mathbf{s})=\bar{C}(\mathbf{s},\pi^{*})+\sum\limits _{\mathbf{s}^{\prime}\in\mathcal{S}}\bar{p}(\mathbf{s}^{\prime}|\mathbf{s},\pi^{*})h(\mathbf{s}^{\prime})$
;

\STATE $h^{'}(\mathbf{s})=h(\mathbf{s})$;

\STATE $h(\mathbf{s})=V(\mathbf{s})-V(\mathbf{s^{\dagger}})$;

\ENDFOR

\UNTIL{$\text{\ensuremath{\mid h(\mathbf{s})-h^{'}(\mathbf{s})\mid}}$<$\bar{\lambda}$}

\STATE \textbf{Output:} The optimal transmission policy $\pi^{*}$.

\end{algorithmic}
\end{algorithm}
\begin{figure*}[!t]
\centering \subfloat[DTMC.]{%
\begin{minipage}[c]{0.7\textwidth}%
\includegraphics[width=1\textwidth]{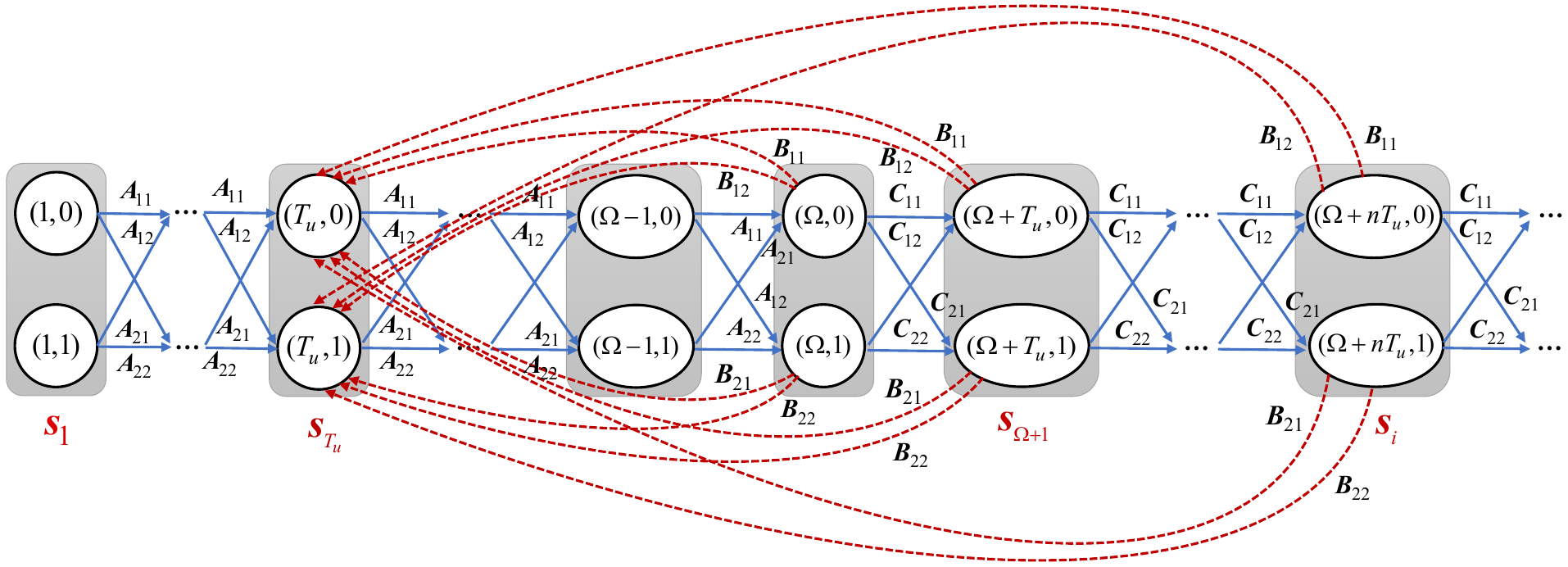}%
\end{minipage}

}\subfloat[Transition probability matrix.]{%
\begin{minipage}[c]{0.28\textwidth}%
\includegraphics[width=1\textwidth]{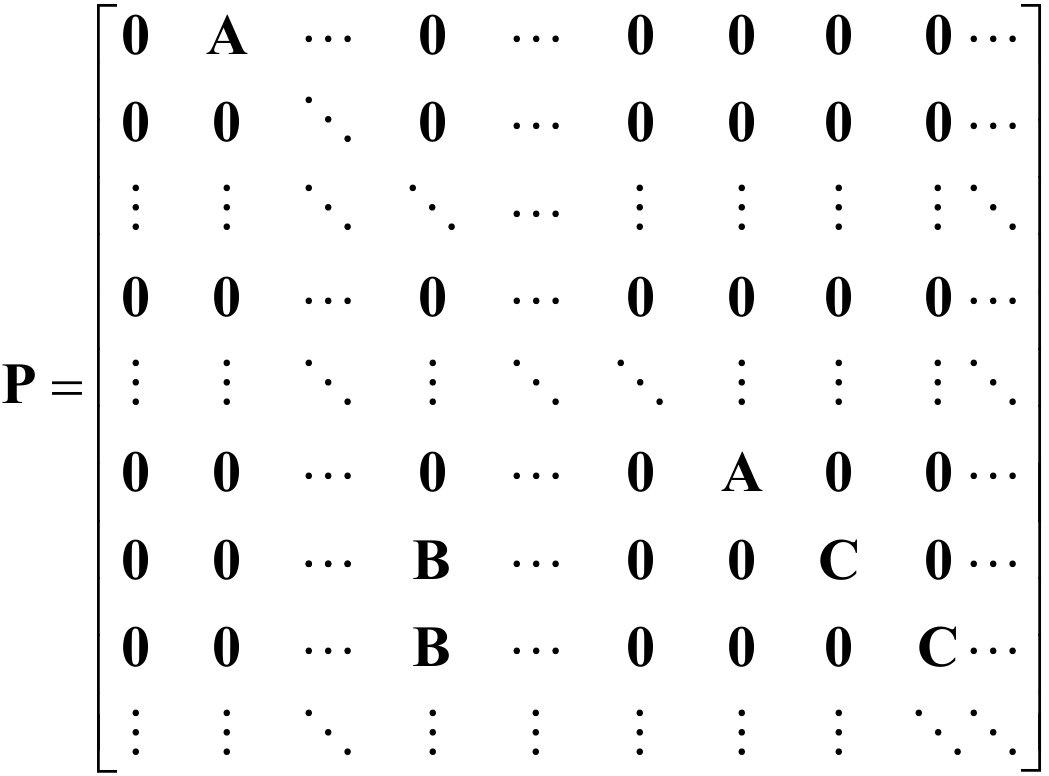}%
\end{minipage}

}\caption{The DTMC and its transition probability matrix.}
\label{Fig:DTMC-tm}

\vspace{-1em}
\end{figure*}

Fig. \ref{Fig:optimal-thresholds-2D-pA} shows that the structure
of the optimal policy for different $p_{A}$. Regardless of the value
of $p_{A}$, the optimal policies corresponding to $F(X)=0$ and $F(X)=1$
always exhibit a threshold structure. This validates Theorem \ref{Th:Theorem1}.
Moreover, when $p_{A}<0.7$ ( i.e., $1-p_{A}>p_{B}$), the threshold
corresponding to $F(X)=0$ is greater than that corresponding to $F(X)=1$.
Conversely, the threshold for $F(X)=0$ is smaller than that for $F(X)=1$.
At $p_{A}=0.7$ ( i.e., $1-p_{A}=p_{B}$), the two thresholds are
equal. This fully conforms to Theorem \ref{Th:Theorem2-1}.

With the threshold structure of the optimal policy, we propose a threshold-based
RVI algorithm, as shown in Algorithm \ref{alg:algorithm1-1}, to 
reduce the computational complexity of the original RVI algorithm.
Specifically, in each iteration, we have to update the optimal actions
for all states by minimizing Eq. (\ref{Eq:RVI-1-1}). However, by
leveraging the threshold property, the optimal action for a state
can be determined immediately if the threshold conditions (lines 5
and 7) are satisfied, thereby avoiding the minimization operation
(line 10). To demonstrate the reduction in computational complexity
achieved by the proposed algorithm, we provide a simulation result.
In the simulation with parameters $T_{u}=10$, $q=0.8$, $p_{u}=1$,
$p_{A}=0.3$, $p_{B}=0.3$, $\hat{\Delta}=500$ and $\bar{\lambda}=0.02$,
the number of minimization operations with the threshold property
(19,056 operations) upon termination of the loop in lines 2--21,
which is approximately 28.02\% of that without using the threshold
property (68,000 operations).

\subsection{A Single-threshold Policy}

In this section, we consider a threshold policy with a single threshold
$\Omega$. Under this single-threshold policy, the state transition
process can be modeled as a discrete time Markov chain (DTMC). To
facilitate the derivation of the steady-state probabilities of the
DTMC, we assume that $\hat{\Delta}$ tends to infinity. Based on this
assumption, the DTMC can be shown in Fig. \ref{Fig:DTMC-tm}(a). To
conveniently represent the transition probability matrix of this DTMC,
we aggregate the states and denote the aggregated state vector sequence
as $\{\boldsymbol{s}_{1},\boldsymbol{s}_{2},\cdots,\boldsymbol{s}_{i},\cdots\}$.
Here, $\boldsymbol{s}_{i}$ is the $i$-th element of the state vector
sequence, defined as follows:
\begin{align}
\boldsymbol{s}_{i}=\begin{cases}
((i,0),(i,1)), & 1\leq i<\Omega\\
((\Omega+(i-\Omega)T_{u},0),(\Omega+(i-\Omega)T_{u},1)), & i\geq\Omega
\end{cases}\label{Eq:State vector}
\end{align}

After the above state aggregation, the transition probability matrix
of the DTMC can be represented in the block form shown in Fig. \ref{Fig:DTMC-tm}(b).
Here, all the blocks are $2\times2$ non-negative matrices. Specifically,
$\mathbf{A}$, $\mathbf{B}$, and $\mathbf{C}$ are given by

\begin{equation}
\begin{array}{c}
\mathbf{A}\triangleq\mathrm{Pr}\{\boldsymbol{s}_{i}\rightarrow\boldsymbol{s}_{i+1}\},\hphantom{}1\leq i<\Omega\\
\mathbf{B}\triangleq\mathrm{Pr}\{\boldsymbol{s}_{i}\rightarrow\boldsymbol{s}_{T_{u}}\},\hphantom{}\hphantom{}\hphantom{}\hphantom{}i\geq\Omega\\
\boldsymbol{\mathbf{C}}\triangleq\mathrm{Pr}\{\boldsymbol{s}_{i}\rightarrow\boldsymbol{s}_{i+1}\},\hphantom{}\hphantom{}\hphantom{}\hphantom{}i\geq\Omega
\end{array}
\end{equation}

where 

{\small{}
\[
\begin{array}{c}
\mathbf{A}=\left[\setlength{\arraycolsep}{3pt}\begin{array}{cc}
1-g & g\\
1-g & g
\end{array}\right],\mathbf{B}=\left[\setlength{\arraycolsep}{3pt}\begin{array}{cc}
\hat{p}_{B}p_{u}^{T_{u}}(1-g) & \hat{p}_{B}p_{u}^{T_{u}}g\\
(1-\hat{p}_{A})p_{u}^{T_{u}}(1-g) & (1-\hat{p}_{A})p_{u}^{T_{u}}g
\end{array}\right]\end{array}
\]
}{\small\par}

{\small{}
\[
\begin{array}{c}
\boldsymbol{\mathbf{C}}=\left[\begin{array}{cc}
(1-\hat{p}_{B}p_{u}^{T_{u}})(1-g) & (1-\hat{p}_{B}p_{u}^{T_{u}})g\\
(1-p_{u}^{T_{u}}+\hat{p}_{A}p_{u}^{T_{u}})(1-g) & (1-p_{u}^{T_{u}}+\hat{p}_{A}p_{u}^{T_{u}})g
\end{array}\right].\end{array}
\]
}{\small\par}

Then, we analyze the stationary distribution of the DTMC. Let $\boldsymbol{\varphi}_{i}$
denote the stationary probability of state $\boldsymbol{s}_{i}$.
According to Fig. \ref{Fig:DTMC-tm}(a), we have the stationary distribution
of the DTMC as follows
\begin{equation}
\begin{cases}
\boldsymbol{\varphi}_{i}=\boldsymbol{0}, & 1\leq i<T_{u}\\
\boldsymbol{\varphi}_{i}=\boldsymbol{\varphi}_{T_{u}}\mathbf{A}^{i-T_{u}}, & T_{u}\leq i<\Omega\\
\boldsymbol{\varphi}_{i}=\boldsymbol{\varphi}_{T_{u}}\mathbf{C}^{i}. & i\geq\Omega
\end{cases}\label{Eq:Stationary}
\end{equation}
Note that $\boldsymbol{\varphi}_{i}$ equals $\boldsymbol{0}$ because
the corresponding state $\boldsymbol{s}_{i}$ is the transient state
when $1\leq i<T_{u}$. The eigenvalues of $\mathbf{C}$ are $0$ and
$(1-\hat{p}_{B}p_{u}^{T_{u}})(1-g)+(1-p_{u}^{T_{u}}+\hat{p}_{A}p_{u}^{T_{u}})g$.
Since $0<\hat{p}_{A}<1$, $0<\hat{p}_{B}<1$, $0<p_{u}\leq1$, and
$0<g\leq1$, the spectral radius is less than $1$, i.e., $\mathrm{sp}(\mathbf{C})<1$.
This implies that as $i$ approaches infinity, the series of matrix
$\mathbf{C}$ converges.

Relying on the normalization equation $\sum_{i=0}^{\infty}\boldsymbol{\varphi}_{i}\boldsymbol{e}=1$,
we can obtain
\begin{align}
\left(\Omega-T_{u}\right)\boldsymbol{\varphi}_{T_{u}}\boldsymbol{e}+\boldsymbol{\varphi}_{T_{u}}\left(\mathbf{I}-\mathbf{C}\right)^{-1}\boldsymbol{e}=1.\label{Eq:Balance equation}
\end{align}
By solving the Eq. (\ref{Eq:Balance equation}), we have
\begin{align}
\boldsymbol{\varphi}_{T_{u}} & {\scriptstyle =}\left[\begin{array}{cc}
\tfrac{1-g}{\left(\Omega-T_{u}\right)+[1-g,g]\left(\mathbf{I}-\mathbf{C}\right)^{-1}\boldsymbol{e}} & \tfrac{g}{\left(\Omega-T_{u}\right)+[1-g,g]\left(\mathbf{I}-\mathbf{C}\right)^{-1}\boldsymbol{e}}\end{array}\right].\label{Eq:Stationary TU}
\end{align}

With the stationary probability $\boldsymbol{\varphi}_{T_{u}}$, the
stationary probabilities of other states can be obtained according
to (\ref{Eq:Stationary}). Given the stationary distribution of the
DTMC under the single-threshold policy, the average cost can be given
by the following lemma.
\begin{lem}
\label{LM:Lemma5} When $\hat{\Delta}$ goes to infinity, for any
threshold $\Omega$, the average cost $J(\Omega)$ of the single-threshold
policy is given by
\begin{align}
J(\Omega) & =\frac{1}{2}(\Omega^{2}-\Omega-T_{u}^{2}+T_{u})\boldsymbol{\varphi}_{T_{u}}\boldsymbol{e}+\Omega\boldsymbol{\varphi}_{T_{u}}\left(\mathbf{I}-\mathbf{C}\right)^{-1}\boldsymbol{\boldsymbol{e}}\nonumber \\
 & +T_{u}\boldsymbol{\varphi}_{T_{u}}\mathbf{C}\left(\mathbf{I}-\mathbf{C}\right)^{-2}\boldsymbol{e}.\label{Eq:total_average_cost}
\end{align}
\end{lem}
\begin{IEEEproof}
\label{PR:proof_Lemma5} For our system, the cost of transmission
is $\Delta$. Based on the (\ref{Eq:State vector}), we have the transmission
cost is $i$ if $1\leq i<\Omega$, the transmission cost is $\Omega+(i-\Omega)T_{u}$
if $i\geq\Omega$. Therefore, the average cost under the single-threshold
policy can be derived as follows:

\noindent 
\begin{align}
 & J(\Omega)\nonumber \\
= & \sum_{i=T_{u}}^{\Omega-1}\boldsymbol{\varphi}_{i}\boldsymbol{e}i+\sum_{i=\Omega}^{\infty}\boldsymbol{\varphi}_{i}\boldsymbol{e}(\Omega+(i-\Omega)T_{u})\nonumber \\
= & \boldsymbol{\varphi}_{T_{u}}\boldsymbol{e}\sum_{i=T_{u}}^{\Omega_{0}-1}i+\Omega\sum_{i=0}^{\infty}\left(\boldsymbol{\varphi}_{T_{u}}\mathbf{C}^{i}\boldsymbol{e}\right)+T_{u}\sum_{i=0}^{\infty}i\left(\boldsymbol{\varphi}_{T_{u}}\mathbf{C}^{i}\boldsymbol{e}\right)\nonumber \\
= & \frac{1}{2}(\Omega^{2}-\Omega-T_{u}^{2}+T_{u})\boldsymbol{\varphi}_{T_{u}}\boldsymbol{e}+\Omega\boldsymbol{\varphi}_{T_{u}}\left(\mathbf{I}-\mathbf{C}\right)^{-1}\boldsymbol{\boldsymbol{e}}\nonumber \\
 & +T_{u}\boldsymbol{\varphi}_{T_{u}}\mathbf{C}\left(\mathbf{I}-\mathbf{C}\right)^{-2}\boldsymbol{e}\text{.}
\end{align}

This concludes the proof of Lemma \ref{LM:Lemma5}.
\end{IEEEproof}
The optimal threshold $\Omega^{*}$ of the single-threshold policy
can be determined by minimizing the average cost $J(\Omega)$. Note
that the determinant of $\mathbf{I}-\mathbf{C}$ is non-zero, which
implies that it is invertible. Due to the complexity of the matrix
differentiation and higher-order matrix calculations involved in deriving
the optimal threshold $\Omega^{*}$ from $J(\Omega)$, it is difficult
to directly obtain a closed-form solution for $\Omega^{*}$. Therefore,
we adopt the Brent algorithm to search for the numerical solution
of the optimal threshold $\Omega^{*}$\cite{Brent2013}. 

With the optimal threshold $\Omega^{*}$, the single-threshold policy
is characterized as follows, regardless of the $F(X)$, $\pi^{*}(\Delta,F(X))=1$
if $\Delta\geq\Omega^{*}$; otherwise, $\pi^{*}(\Delta,F(X))=0$.
Compared to the threshold-based RVI introduced in the previous section,
the single-threshold policy has a superior convergence speed. In the
simulation with parameters $T_{u}=10$, $q=0.8$, $p_{u}=1$, $p_{A}=0.3$,
$p_{B}=0.3$, $\hat{\Delta}=500$ and $\bar{\lambda}=0.02$, the convergence
time for the numerical iteration of the approximately optimal policy
is $1.334$ ms, which is much shorter than the $384.482$ ms required
by the threshold-based RVI.

\section{Experiments and Simulations}

\label{Sec:Section 5}In this section, we first validate the feasibility
and effectiveness of the optimal policy for the dynamic transmission
problem through experiments. Then, numerical simulations are conducted
to evaluate the performance of the optimal policy.

\subsection{Experimental Results}

We evaluate the proposed policies using the CIFAR-10 in the experiments.
The CIFAR-10 consists of $60,000$ $32\times32$ RGB images, including
$50,000$ training images and $10,000$ test images. The original
labels range from $\{0,\cdots,9\}$, corresponding to the objects
depicted in each image. We repartition the dataset's labels with binary
labels $\{0,1\}$ to indicate whether the image contains a vehicle.
Specifically, images labeled as $1$ contain a vehicle, while those
labeled as $0$ do not. For the relabeled dataset, we train four networks
with different classification accuracies, as shown in Table \ref{Ta:Comp_acc}.

To assess the effectiveness of the proposed policies in our remote
monitoring system, we assume that the target of the remote monitoring
is vehicles, i.e., the images labeled as $1$ in the CIFAR-10 are
the target transmission data for the system. Then, by selecting different
pre-trained classification networks as the system's processor, the
misidentification probabilities $p_{A}$ and $p_{B}$ can be adjusted.
At the beginning of each decision epoch (i.e., time step) in the inference
phase, the sensor selects an image labeled as $1$ from the CIFAR-10
test set with probability $q$. After the inference process reaches
the set time limit, the performance of the proposed policies is evaluated
by comparing their average TAoIs.

\begin{table}[!h]
\centering \caption{Compare pre-identification accuracy on CIFAR-10}
\begin{tabular}{ccc}
\toprule 
\multirow{2}{*}{Network} & Test Accuracy of Vehicles & Test Accuracy of Animals\tabularnewline
 & $1-p_{A}$ & $1-p_{B}$\tabularnewline
\midrule 
LeNet \cite{Lecun1998} & $61.63\%$ & $59.76\%$\tabularnewline
AlexNet \cite{Krizhevsky2012} & $78.76\%$ & $75.67\%$\tabularnewline
VGG-16 \cite{Simonyan2015} & $84.53\%$ & $82.60\%$\tabularnewline
ResNet-18 \cite{He2016} & $95.17\%$ & $93.52\%$\tabularnewline
\bottomrule
\end{tabular}\label{Ta:Comp_acc}
\end{table}

\begin{figure}[!t]
\centering \includegraphics[width=0.45\textwidth]{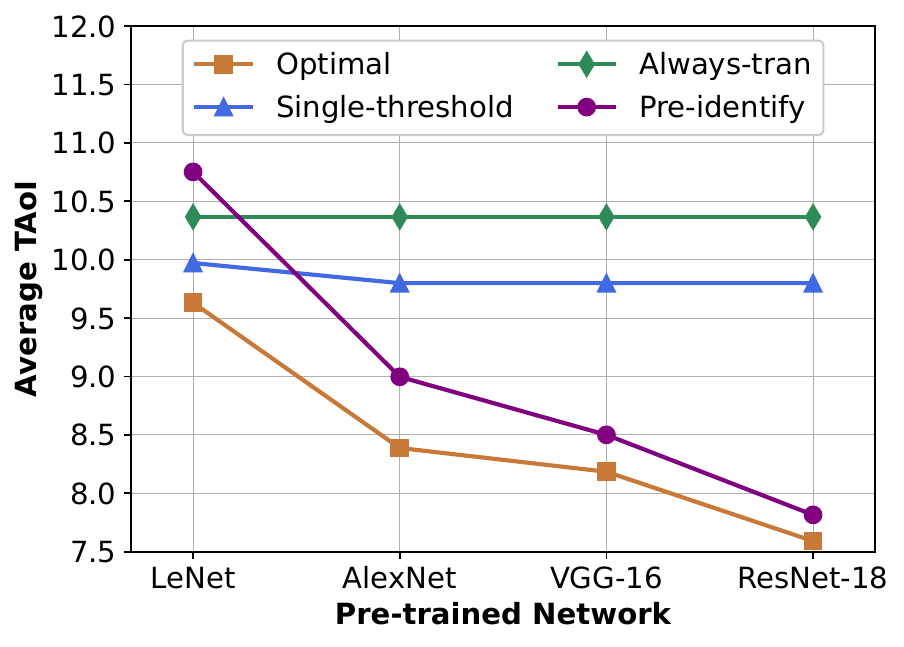}\caption{Average TAoI for different pre-identification networks ($T_{u}=4$,
$q=0.5$, $p_{u}=0.98$).}

\label{Fig:TAoI_network}\vspace{-1em}
\end{figure}

We evaluate two proposed policies: the optimal policy, derived using
a threshold-based RVI approach, and the single-threshold policy, which
employs a fixed threshold for transmission decisions. These are compared
with two baseline policies: the always transmit policy and the pre-identification
based policy. The always transmit policy serves as a benchmark, initiating
a new transmission immediately after the previous one finishes, irrespective
of the current state (i.e., TAoI or the pre-identification result).
Conversely, the pre-identification based policy operates strictly
on the pre-identification outcome. Specifically, the receiver requests
data only when the pre-identification result is 1; otherwise, no transmission
occurs.

Fig. \ref{Fig:TAoI_network} shows the average TAoI for  different
pre-identification networks (LeNet, AlexNet, VGG-16, ResNet-18), which
represent increasing pre-identification accuracy. A consistent observation
is that the optimal policy achieves a significantly lower average
TAoI compared to all other policies, regardless of the pre-identification
network employed. The performance of the single-threshold policy is
better than the pre-identification policy when network accuracy is
lower (e.g., with LeNet). However, as network accuracy improves (from
AlexNet to ResNet-18), the pre-identification policy demonstrates
superior performance over the single-threshold policy. In contrast,
the always transmit policy maintains a constant average TAoI across
all pre-identification networks, as its operation is independent of
the processor's pre-identification capabilities.

\begin{figure}[!t]
\centering \includegraphics[width=0.45\textwidth]{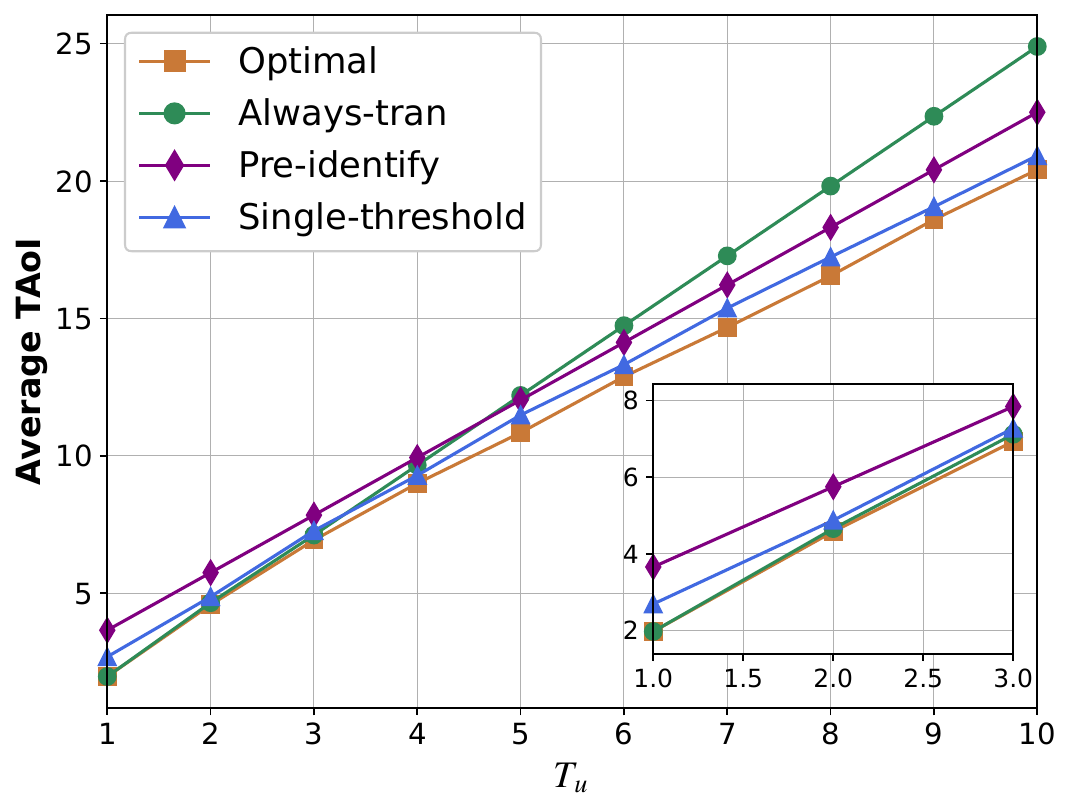}\caption{Average TAoI versus $T_{u}$ ($q=0.5$, $p_{A}=0.4$, $p_{B}=0.4$,
$p_{u}=1$).}

\vspace{-1em}
\label{Fig:TAoI_TU}
\end{figure}

\begin{figure}[!t]
\centering \includegraphics[width=0.45\textwidth]{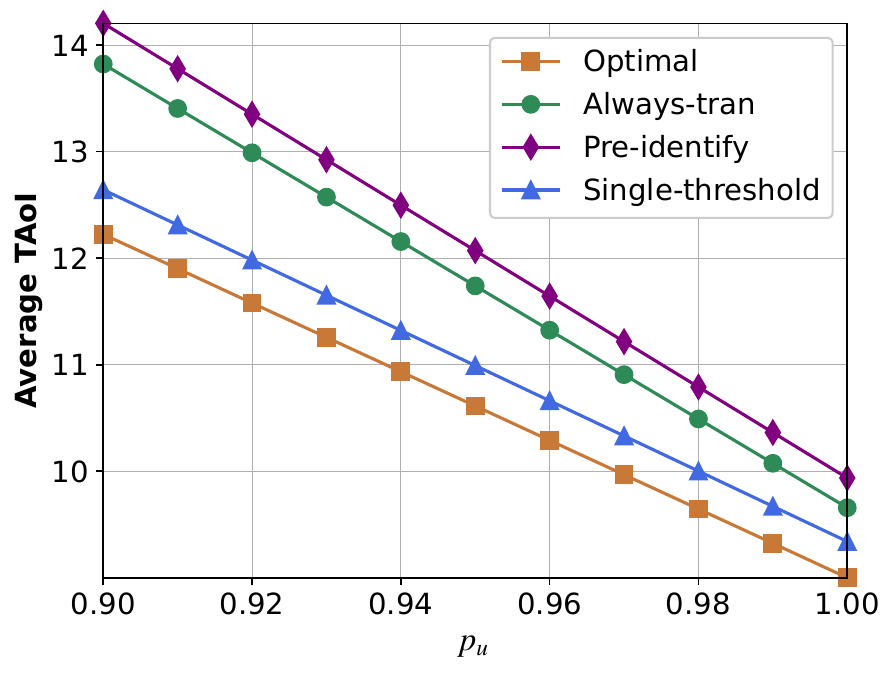}\caption{Average TAoI versus $p_{u}$ ($T_{u}=4$, $q=0.5$, $p_{A}=0.4$,
$p_{B}=0.4$).}

\label{Fig:TAoI_Tran}\vspace{-1em}
\end{figure}

\begin{figure}[!t]
\centering \includegraphics[width=0.45\textwidth]{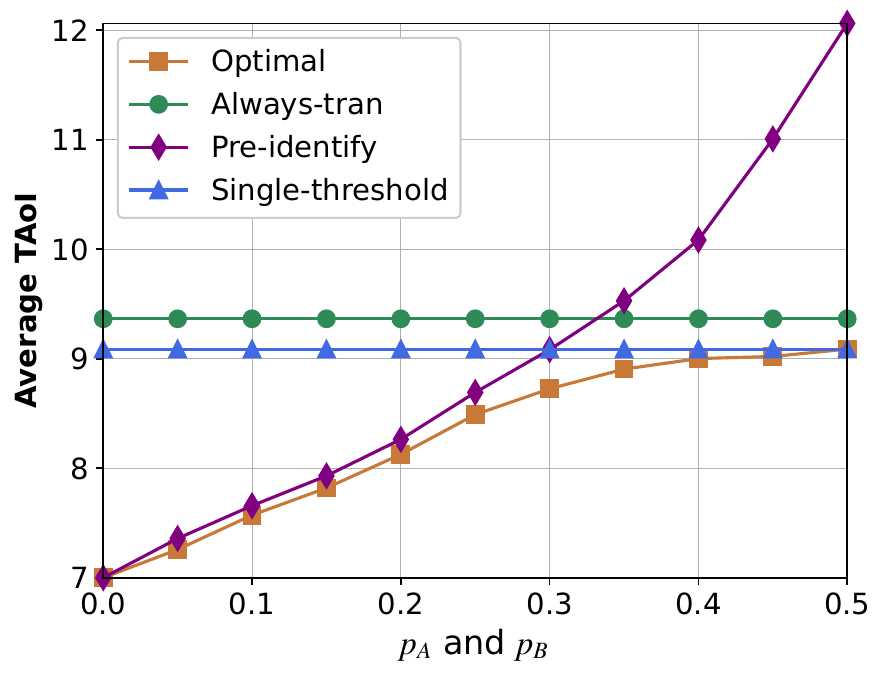}\caption{Average TAoI versus $p_{A}$ and $p_{B}$ ($T_{u}=4$, $q=0.5$, $p_{u}=1$).}

\label{Fig:TAoI_PAB}
\end{figure}

\begin{figure}[!t]
\centering \includegraphics[width=0.45\textwidth]{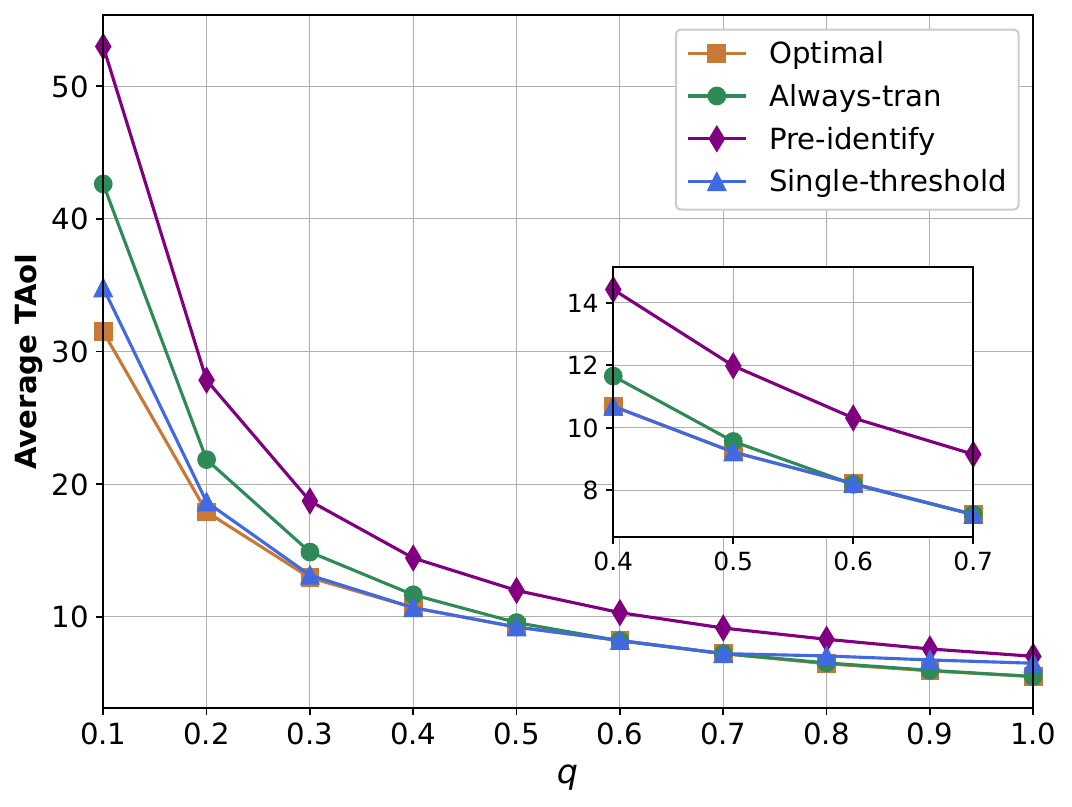}\caption{Average TAoI versus $q$ ($T_{u}=4$, $p_{A}=0.5$, $p_{B}=0.5$,
$p_{u}=1$).}

\label{Fig:TAoI_q}\vspace{-1.5em}
\end{figure}

\subsection{Simulation Results}

\begin{figure*}[!t]
\centering \subfloat[$q=0.1$, $p_{B}=0$]{\includegraphics[width=0.24\textwidth]{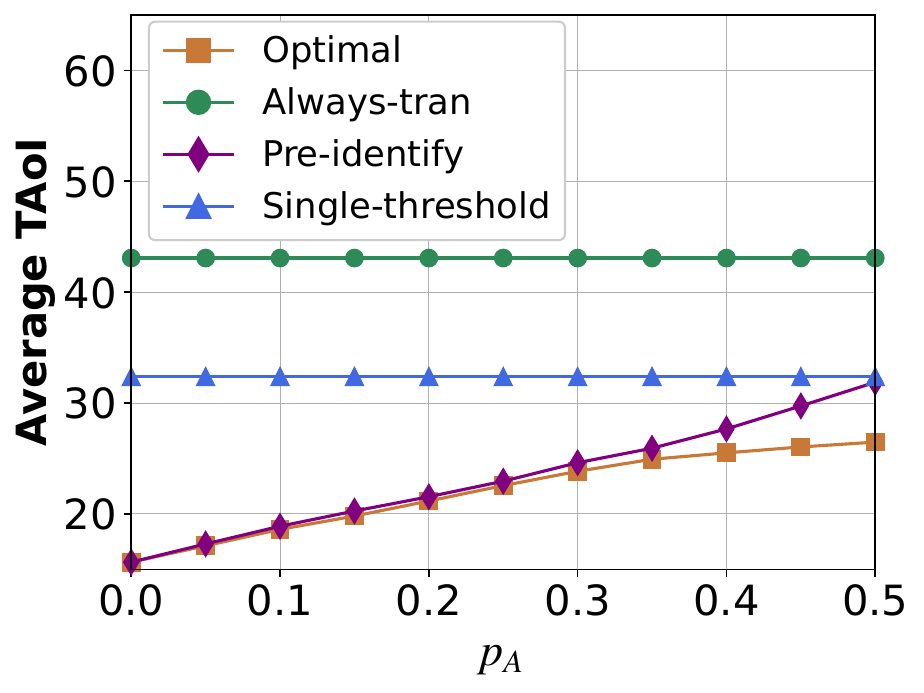}

}\subfloat[$q=0.1$, $p_{A}=0$]{\includegraphics[width=0.24\textwidth]{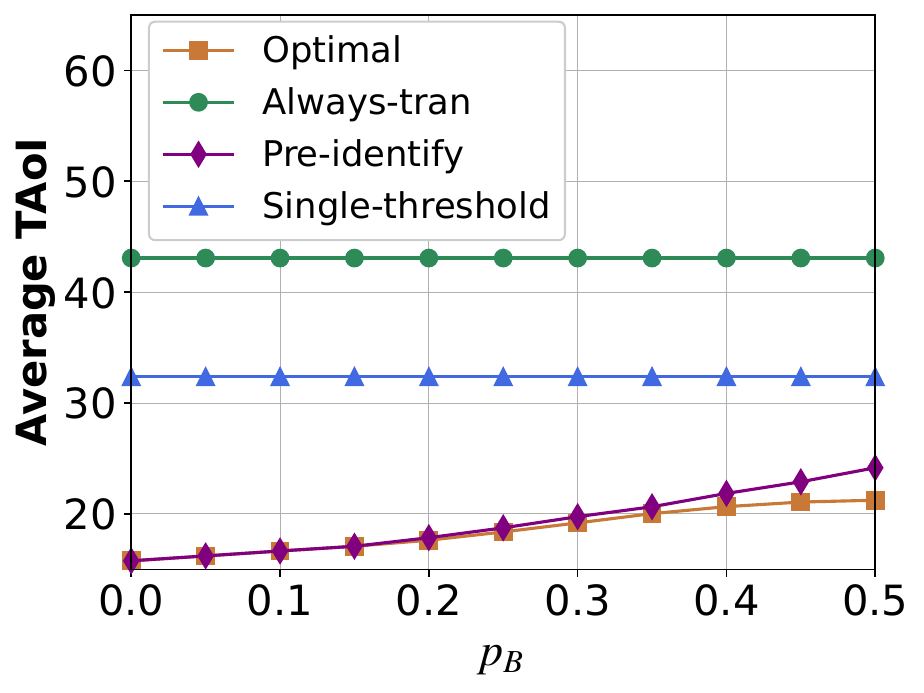}

}\subfloat[$q=0.9$, $p_{B}=0$]{\includegraphics[width=0.24\textwidth]{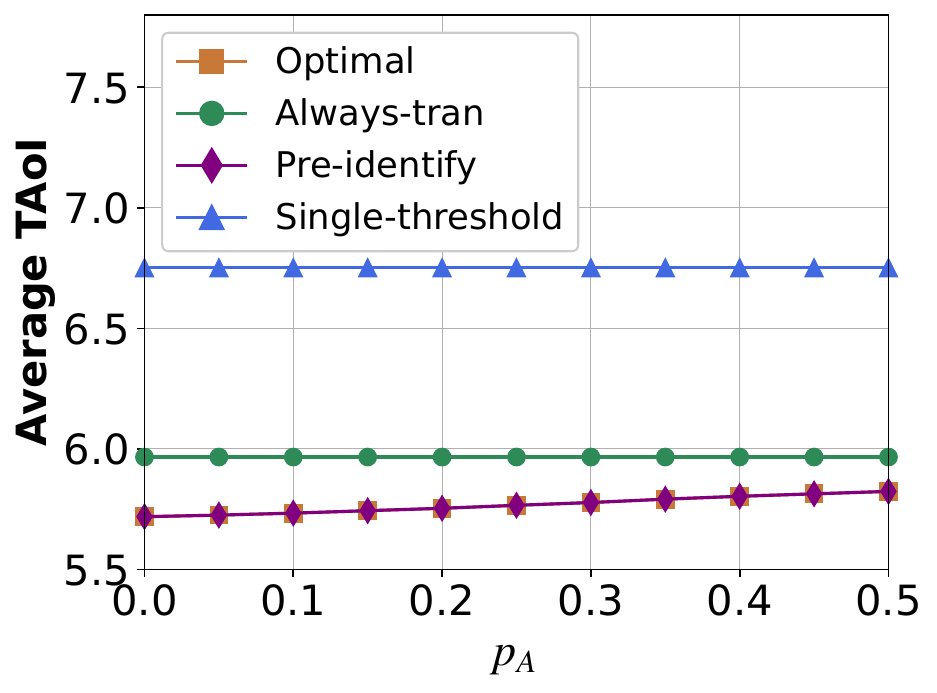}

}\subfloat[$q=0.9$, $p_{A}=0$]{\includegraphics[width=0.24\textwidth]{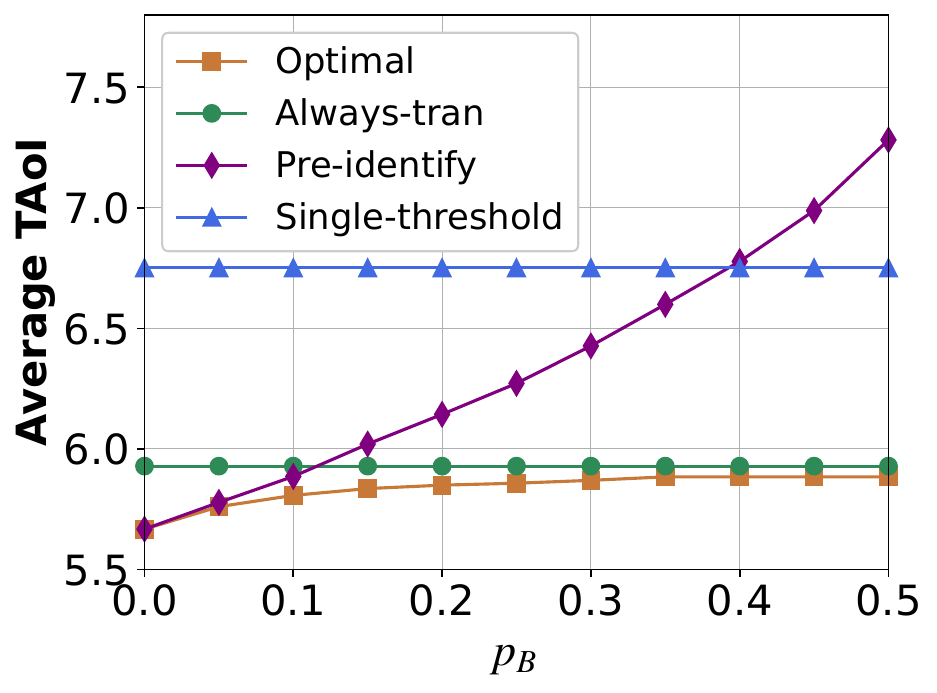}

}\caption{Average TAoI versus $p_{A}$ or $p_{B}$ ($T_{u}=4,p_{u}=1$).}
\label{Fig:TAoI_PAPB}
\end{figure*}

In Fig. \ref{Fig:TAoI_TU}, the average TAoI of the two proposed 
and  baseline policies are compared with respect to the transmission
cost $T_{u}$. A primary observation is that as $T_{u}$ increases,
the advantages of our proposed two policies become increasingly evident.
Throughout the range of $T_{u}$, the single-threshold policy consistently
remains slightly inferior to the optimal policy. This difference arises
because the single-threshold policy neglects the instantaneous pre-identification
results.  Furthermore, at very low $T_{u}$ values, the optimal policy
converges with the always transmit policy. Specifically, when $T_{u}=1$,
the optimal policy exactly coincides  with the always transmit policy.
This occurs because, when the transmission cost equals $1$, it precisely
matches the cost of not transmitting, leading the optimal policy to
always choose transmission due to the potential to capture the target.
 Conversely, when $T_{u}$ becomes large ($T_{u}>4$), the always-transmit
policy exhibits the worst performance,  as the increasing cost associated
with transmission failures in the always-transmit policy becomes a
significant disadvantage.

In Fig. \ref{Fig:TAoI_Tran}, the average TAoI of the optimal policy
and the two policies are compared with respect to the transmission
success probability $p_{u}$. Notably, our analysis focuses on higher
$p_{u}$ values (0.9 to 1) due to the exponential decrease in overall
transmission success probability $p_{u}^{T_{u}}$ with increasing
$T_{u}$.  It can be observed from Fig. \ref{Fig:TAoI_Tran} that
as $p_{u}$ increases, the average TAoIs of all policies consistently
decreases. This is intuitive, as a higher transmission success probability
directly corresponds to a higher success rate for monitoring tasks,
leading to fresher data. Moreover, as  $p_{u}$ decreases, the performance
advantage of the two proposed policies becomes increasingly evident.
This occurs because the two baseline policies do not account for variations
in $p_{u}$ when determining their policies, whereas the two proposed
policies  explicitly consider $p_{u}$ in their decision-making. This
allows them to adapt more effectively to less reliable transmission
environments.

In Fig. \ref{Fig:TAoI_PAB}, we compare the average TAoI of all policies
with respect to the misidentification probabilities $p_{A}$ and $p_{B}$.
First, as observed in Fig. \ref{Fig:TAoI_PAB}, the average TAoI of
the always-transmit policy and the single-threshold policy remains
constant across the entire range of $p_{A}$ and $p_{B}$.  The always-transmit
policy is inherently unaffected by misidentification probabilities.
For the single-threshold policy, its constant TAoI stems from its
insensitivity to variations  in $p_{A}$ and $p_{B}$ under the current
parameter settings. Specifically, according to Eq. (\ref{Eq:total_average_cost}),
the threshold $\Omega$ of the single-threshold policy primarily depends
on the transmission cost $T_{u}$. When $T_{u}$ is relatively small
(e.g., set as $4$ in this simulation), variations in $p_{A}$ and
$p_{B}$ have negligible effects on the threshold, keeping it consistently
at $6$. In contrast, the pre-identification based policy performs
well when $p_{A}$ and $p_{B}$ are small, and achieves the same performance
as the optimal policy when the misidentification probabilities $p_{A}$
and $p_{B}$ are zero. This is because the pre-identification based
policy relies solely on the pre-identification result, and when the
misidentification probability is low, its performance naturally improves.
Finally, when the misidentification probabilities $p_{A}$ and $p_{B}$
are both equal to $0.5$, the average TAoI of the single-threshold
policy is the same as that of the optimal policy. This equivalence
arises because, under these conditions, the optimal policy's thresholds
$\Omega_{0}$ and $\Omega_{1}$ both become 6, precisely matching
the threshold $\Omega$ of the single-threshold policy.

In Fig. \ref{Fig:TAoI_q}, we compare the average TAoI of the two
proposed policies and two baseline policies with the varying probability
 of target appearance. As anticipated, the TAoI of all policies decreases
with increasing $q$, which is attributed to the higher frequency
of target appearance leading to an increased probability of successful
monitoring.  When the probability of target image appearance $q$
reaches $1$, the optimal policy and the always transmit policy converge
to the same average TAoI.  This indicates that when the target is
guaranteed to appear, continuous transmission becomes the optimal
strategy. In contrast, the pre-identification policy generally results
in a higher TAoI compared to the proposed policies across the entire
$q$ range. Furthermore, a point of near-identical performance between
the optimal policy and the single-threshold policy is observed around
$q=0.5$. As shown in the figure, particularly when $q$ is close
to $0.5$ (e.g., $q=0.4,0.6,0.7$), their average TAoIs are nearly
identical. This behavior aligns with the case where $p_{A}=p_{B}=0.5$
in Fig. \ref{Fig:TAoI_PAB}. According to Theorem \ref{Th:Theorem2-1},
when $p_{A}=1-p_{B}$ (which holds in this figure), the thresholds
$\Omega_{0}$ and $\Omega_{1}$ generated by the threshold-based RVI
become equal. In this scenario, the threshold $\Omega$ of the single-threshold
policy also aligns with $\Omega_{0}$ and $\Omega_{1}$, making the
single-threshold policy effectively equivalent to the optimal policy.

Fig. \ref{Fig:TAoI_PAPB} further illustrates the individual impacts
of misidentification probabilities $p_{A}$ and $p_{B}$ on each policy's
performance. The figure presents four distinct scenarios, all with
a transmission success probability $p_{u}=1$: (a) varying $p_{A}$
with $q=0.1,p_{B}=0$; (b) varying $p_{B}$ with $q=0.1,p_{A}=0$;
(c) varying $p_{A}$ with $q=0.9,p_{B}=0$; and (d) varying $p_{B}$
with $q=0.9,p_{A}=0$. Consistent with observations from Fig. \ref{Fig:TAoI_PAB},
both the always-transmit policy and the single-threshold policy remain
constant across all subfigures, as their design is unaffected by changes
in $p_{A}$ or $p_{B}$. By comparing Fig. \ref{Fig:TAoI_PAPB}(a)
with Fig. \ref{Fig:TAoI_PAPB}(b), it is evident that when $q$ is
small, the pre-identification based policy's TAoI is more sensitive
to variations in $p_{A}$ than $p_{B}$. Conversely, for a large $q$,
comparing Fig. \ref{Fig:TAoI_PAPB}(c) with Fig. \ref{Fig:TAoI_PAPB}(d)
reveals greater sensitivity to $p_{B}$. This behavior is directly
linked to the policy's reliance on pre-identification accuracy: $p_{A}$
has a greater impact when actual targets are rare, while $p_{B}$
has a more pronounced effect when targets are frequent. The optimal
policy, in contrast, demonstrates remarkable stability, with its performance
remaining relatively robust to variations in both $p_{A}$ and $p_{B}$,
irrespective of $q$.

\section{Conclusions}

\label{Sec:Section 6}

In this paper, we introduced a new task-oriented communication metric
named TAoI, which directly measures the relevance of information content
to the system's task, thereby enhancing the system's decision-making
utility. Our primary objective was to minimize TAoI by exploring the
optimal transmission policy for remote monitoring systems with pre-identification.
To this end, we modeled the dynamic transmission problem as an SMDP
over an infinite time horizon and converted it into an equivalent
MDP with uniform time steps. Our theoretical analysis revealed that
the optimal transmission policy is a threshold policy.  Based on this
insight, we proposed a low-complexity, threshold-structured relative
value iteration algorithm to effectively obtain this optimal policy.
Furthermore, a simpler single-threshold policy was introduced, which,
while exhibiting a slight performance degradation, offers the significant
advantage of faster convergence. Finally, comprehensive simulation
results showed that the optimal transmission policy significantly
outperforms baseline policies.

\bibliographystyle{IEEEtran}
\bibliography{IEEEabrv,TAoI_Jour_Ref}

\end{document}